\documentclass[10pt,journal,twocolumn,twoside]{IEEEtran} 

\usepackage{graphicx}
\usepackage{epstopdf}
\usepackage{float}
\usepackage{algorithmic}
\usepackage{array}
\usepackage{amsmath}
\usepackage{amssymb}
\usepackage{mdwmath}
\usepackage{mdwtab}
\usepackage{eqparbox}
\usepackage{stfloats}
\usepackage{fixltx2e}
\usepackage{cases} 

\usepackage{flushend}

\usepackage{xcolor}
\usepackage{makecell}

\usepackage[boxed,ruled,commentsnumbered]{algorithm2e}
\usepackage{url}
\usepackage{cite}
\ifCLASSOPTIONcompsoc
\usepackage[caption=false,font=normalsize,labelfont=sf,textfont=sf]{subfig}
\else
\usepackage[caption=false,font=footnotesize]{subfig}
\fi
\allowdisplaybreaks[4]

\makeatletter
\renewcommand*{\@opargbegintheorem}[3]{\trivlist
      \item[\hskip \labelsep{\bfseries #1\ #2}] \textbf{(#3):}\ }
\makeatother

\begin{document}

\title
{Vehicular Behavior-Aware Beamforming Design for Integrated Sensing and Communication Systems}
\author{Dingyan Cong,~\IEEEmembership{Student Member, IEEE,} Shuaishuai Guo,~\IEEEmembership{Senior Member, IEEE,}  Shuping Dang,~\IEEEmembership{Member, IEEE,} and Haixia Zhang,~\IEEEmembership{Senior Member, IEEE}

\thanks{Dingyan Cong, Shuaishuai Guo, and Haixia Zhang are with School of Control Science and Engineering and also  with Shandong Provincial Key Laboratory of Wireless Communication Technologies, Shandong University, Jinan 250061, China (e-mail: dingyan.cong@mail.sdu.edu.cn, shuaishuai\textunderscore guo@sdu.edu.cn, haixia.zhang@sdu.edu.cn).}
\thanks{Shuping Dang is with the Department of Electrical and Electronic Engineering, University of Bristol, Bristol BS8 1UB, UK (e-mail: shuping.dang@bristol.ac.uk).}
   }
\maketitle


\begin{abstract} 
Communication and sensing are two important features of connected and autonomous vehicles (CAVs). In traditional vehicle-mounted devices, communication and sensing modules exist but in an isolated way, resulting in a waste of hardware resources and wireless spectrum. In this paper, to cope with the above inefficiency, we propose a vehicular behavior-aware integrated sensing and communication (VBA-ISAC) beamforming design for the vehicle-mounted transmitter with multiple antennas. In this work, beams are steered based on vehicular behaviors to assist driving and meanwhile provide spectral-efficient uplink data services with the help of a roadside unit (RSU). Specifically, we first predict the area of interest (AoI) to be sensed based on the vehicles' trajectories. Then, we formulate a VBA-ISAC beamforming design problem to sense the AoI while maximizing the spectral efficiency of uplink communications, where a trade-off factor is introduced to balance the communication and sensing performance. A semi-definite relaxation-based beampattern mismatch minimization (SDR-BMM) algorithm is proposed to solve the formulated problem. To reduce the hardware cost and power consumption, we further improve the proposed VBA-ISAC beamforming design by introducing the hybrid analog-digital (HAD) structure.  
Numerical results verify the effectiveness of VBA-ISAC scheme and show that the proposed beamforming design outperforms the benchmarks in both spectral efficiency and radar beampattern.

\end{abstract}


\section{Introduction} 

\IEEEPARstart{C}{onnected}  and autonomous vehicles (CAVs) are the next frontier of the automotive revolution and the key of innovation in the next-generation intelligent transportation systems (ITS) \cite{joshua2018survey}. In future ITS, CAVs will be not only a means of smart transportation but also a service platform similar to a mobile phone, providing passengers with fast and secure data services. To achieve this vision, a vehicle needs to have two typical capabilities: sensing and communications \cite{bei2022optimal}. On the one hand, to support safe and autonomous driving, the vehicle needs to sense the environment through radars to obtain environmental information, such as the distance to the vehicle ahead and its speed. On the other hand, the vehicle needs to communicate with other vehicles, passengers, and infrastructures through vehicle-mounted transceivers. In existing vehicle-mounted devices, communication and sensing functional modules exist but in an isolated way, resulting in a waste of hardware resources and wireless spectrum \cite{liu2020joint}. To enable the efficient use of the spectrum resources and reduce the hardware cost, the integrated sensing and communication (ISAC) technology was put forward, where two functions of communication and sensing are integrated into the same device \cite{fan2022integrated}. 
For vehicular-mounted ISAC devices with multi-antennas, how to design the beamformer is an important and intricate task, since the beamforming design has to meet two kinds of quality of service (QoS) and balance the communication and sensing performance. In this paper, we investigate this crucial research problem by being aware of vehicular behaviors.

\subsection{Prior Works}
The investigation on the ISAC beamforming design is quickly gaining traction in the communication and signal processing community due to the prospect of integrating dual functionalities of radar sensing and communications.
Previous myriad ISAC designs can be mainly classified into three categories: communication-centric ISAC design, sensing-centric ISAC design, and balanced ISAC design.

\subsubsection{Communication-centric ISAC design} In such systems, sensing comes into play with the assistance of communications. Sensing is adopted to mitigate interference and improve spectral efficiency for communications. Biswas \emph{et al} in \cite{sudip2020design} proposed a multiple-input multiple-output (MIMO) radar to MIMO communication systems to tackle imperfect channel estimation and hardware impairments while improving the QoS for cellular users. Based on the jointly phased arrays, Feng and Huang in \cite{Feng2020precoding} designed the beamformer by jointly optimizing the interference between communications and radar sensing to improve the received signal-to-noise ratio (SNR).  Liu \emph{et al} in \cite{xiang2020joint} proposed to simultaneously transmit the integrated radar waveform and constellation symbols while ensuring the SNR of each user is above a preset threshold. To reduce the pilot overhead for beam alignment and channel estimation,\cite{mas2017robust,nuria2016radar} introduced a millimeter-wave (mmWave) radar in the base station.  Huang \emph{et al} in \cite{huang2021MIMO} proposed a deep-learning-enabled MIMO-radar-assisted channel estimation scheme. They designed the transmission frame structure of the combined radar sensing module and communication module while estimating the angle of departure and the angle of arrival and establishing a stable communication link. Shen \emph{et al} in \cite{shen2019channel} proposed to use the orthogonal time frequency space (OTFS) modulation waveform to  sense the delay and the Doppler shift of wireless channels. Since the channel state information (CSI) is obtained with low pilot overhead,  the spectral efficiency of the communication system is significantly improved. Shaham \emph{et al} in \cite{sina2019fast} have achieved a similar research goal of channel estimation by using sensing to assist communications.

\subsubsection{Sensing-centric ISAC design} In this category, communication is an auxiliary function to assist sensing, enhancing the sensing performance, including improving sensing accuracy and developing sensing functionality in existing communication systems. Barneto \emph{et al} in \cite{carlos2020beamforming} and Damith \emph{et al} in \cite{sahan2021joint} optimized the transmitting and receiving beamformers to enable multi-beam sensing.  Keskin \emph{et al} in \cite{MUSA2021limited} studied the time-frequency waveform design of radar and communication systems. And they focused on the research of the radar optimal waveform design that minimizes the Cram\'{e}r-Rao bound on the delay-Doppler estimation in the delay-Doppler ambiguity domain, aiming at improving the radar sensing accuracy and resolution. Takahara \emph{et al} in \cite{takahara2012study} proposed a communication-assisted ultra-wideband radar system to achieve high-precision ranging and positioning.  Wymeersch \emph{et al} in \cite{henk20175g} proposed to combine cellular networks with the existing vehicle positioning and map systems.  
Besides, early scholars working on ISAC paid more attention to increasing sensing in communication systems. The authors of \cite{daniels2018forward, kumari2018ieee} extended the WiFi technology to radar systems. Daniels \emph{et al} in \cite{daniels2018forward} proposed a method to determine the average normalized channel energy from the frequency-domain channel estimation and modeled it as a simple sinusoidal of the target distance so as to achieve the closest target distance estimation. Kumari \emph{et al} in \cite{kumari2018ieee} had done similar work by applying a radar in WiFi systems. They developed single-frame and multi-frame radar receiver algorithms for target detection as well as distance and speed estimations for single-target and multi-target scenarios.

\subsubsection{Balanced ISAC design} Unlike the communication-centric ISAC design and sensing-centric ISAC design, in the balanced ISAC design, sensing and communication functions are of equal importance and both play crucial roles. The authors of \cite{cheng2021wideband,tang2020waveform,sayed2021mult,liu2018toward,liu2019hybrid,dong2022vpaitmin} focused on the flexible performance trade-off between communications and sensing. Specifically, Cheng \emph{et al} in \cite{cheng2021wideband} maximized the communication rate while having good sensing beampattern characteristics under power constraints. Tang \emph{et al} in \cite{tang2020waveform}  used a dual-function MIMO array, which can match a desired transmit beampattern for radar sensing and to communicate with multiple users simultaneously. Dokhanchi \emph{et al} in \cite{sayed2021mult} dedicated to the beamforming design of the Internet of Vehicles (IoV) system. In the IoV system, the transmitter communicates with multiple vehicles, and in the meantime, the radar detects multiple targets. The beam is designed to maximize the communication rate under a constraint on radar detection performance. Liu \emph{et al} in \cite{liu2018toward} considered the full-digital beamforming design for MIMO dual-function radar and communication system by weighting optimization for the flexible trade-off between radar sensing and communication. 
\cite{liu2019hybrid} and \cite{dong2022vpaitmin} exploited the ISAC beamforming design with a hybrid radio frequency (RF) structure in vehicle-to-everything (V2X) scenarios, minimizing the weighted sum of communication and radar sensing beamforming errors.

The aforementioned works have well investigated the realization of ISAC beamforming designs. In particular, in \cite{liu2019hybrid,dong2022vpaitmin}, they consider the ISAC beamforming design for V2X scenarios. Their works \cite{liu2018toward,liu2019hybrid,dong2022vpaitmin} brings important insights of beamforming design for ISAC systems, but it is not appropriate to apply directly on V2X systems. 
The most significant problem is that the targets to be sensed are assumed to be fixed in their designs. 
In more detail, their schemes do not take vehicle behavior into account when determining the sensing area. When the vehicle is moving, their schemes can't get the surrounding environment information in advance, and cannot accurately set the required pointing angles. Moreover, their beamforming design also cannot obtain the required sensing distance at each pointing angle. In the case of limited energy of vehicular-mounted ISAC device, it can be hardly to accurately cover the area that needs to be sensed.

Therefore, such an assumption in ISAC design is not suitable for vehicular networks with high mobility. When CAVs are moving, they need to be aware of surrounding environments in advance. 
The ISAC beamforming design for vehicle-mounted transmitters should thereby take the mobility and behaviors of vehicles into consideration to perform predictive sensing and simultaneous communications. In this regard, different from myriad previous works, we propose a vehicular behavior-aware ISAC (VBA-ISAC) beamforming design for vehicle-mounted multi-antenna transmitters in this paper. Specifically, beams are designed to steer based on vehicles' behavior to predicatively sense the area of interest (AoI) and meanwhile provide spectrum-efficient uplink data services with the help of roadside unit (RSU).

\subsection{Contributions}
To be clear, we summarize the technical contributions of this paper in detail as follows: 

\begin{itemize}
\item We propose a VBA-ISAC beamforming scheme for multi-antenna vehicle-mounted transmitters to simultaneously provide communication and predictive sensing capabilities to vehicles. The proposed scheme is capable of predicting the AoI according to the real-time behavior of the vehicle via in-vehicle sensors. And the optimal radar beamformer is designed to exactly cover the AoI.
\item We formulate an optimization problem to maximize the spectral efficiency of communications while minimizing the beampattern mismatch to optimal radar beamformer, where a trade-off factor is introduced to balance the communication and sensing performance. To solve the formulated optimization problem, we also propose a semi-definite relaxation-based beampattern mismatch minimization (SDR-BMM) algorithm. 
\item To reduce the power consumption and hardware cost of multiple radio frequency (RF) chains supporting multi-antenna transmitters, we further investigate the VBA-ISAC beamforming design with hybrid RF chains structure. 
\end{itemize}
We analyze and simulate the sensing performance and communication performance of the proposed designs in this paper, which demonstrate and quantify clear performance advantages over the benchmarks.

\subsection{Organization}
The remainder of this paper is organized as follows. We present the system model in Section~\ref{II}. In Section~\ref{III}, we formulate the VBA-ISAC beamforming design problem and propose the solution to the formulation. We study the VBA-ISAC beamforming designs with the hybrid structure in Section~\ref{IV}. The numerical results are presented and discussed in Section~\ref{V}, and we finally conclude this paper in Section~\ref{VI}.

\subsection{Notations}
In this paper, the notations are defined and used in the following manner. $\mathbf{A}$ and $\mathbf{a}$ stand for a matrix and a column vector, respectively; $\mathbf{A}_{i,j}$ is the entry on the $i$th row and $j$th column of matrix $\mathbf{A}$; $\mathbf{(\cdot)}^*$, $\mathbf{(\cdot)}^T$ and $\mathbf{(\cdot)}^H$ stand for the conjugate, transpose, and conjugate transpose operations of the matrix or a vector enclosed; The determinant and Frobenius norm of a matrix are represented by $\det(\mathbf{A})$ and $\|\mathbf{A}\|_F$, respectively; $\mathbf{A}^{-1}$ and $\mathbf{A}^{\dag}$ are the inverse and Moore-Penrose pseudo inverse of matrix $\mathbf{A}$; $\mathrm{vec}(\mathbf{\cdot})$ indicates vectorization of the matrix enclosed; Expectation and the real part of the complex variable enclosed are denoted by $\mathbb{E}[\cdot]$ and $\Re[\cdot]$; Hadamard and Kronecker products between two matrices are represented by $\circ$ and $\otimes$, respectively.

\section{System Model} \label{II}

\begin{figure}
  \centering
  \includegraphics[width=0.40\textwidth]{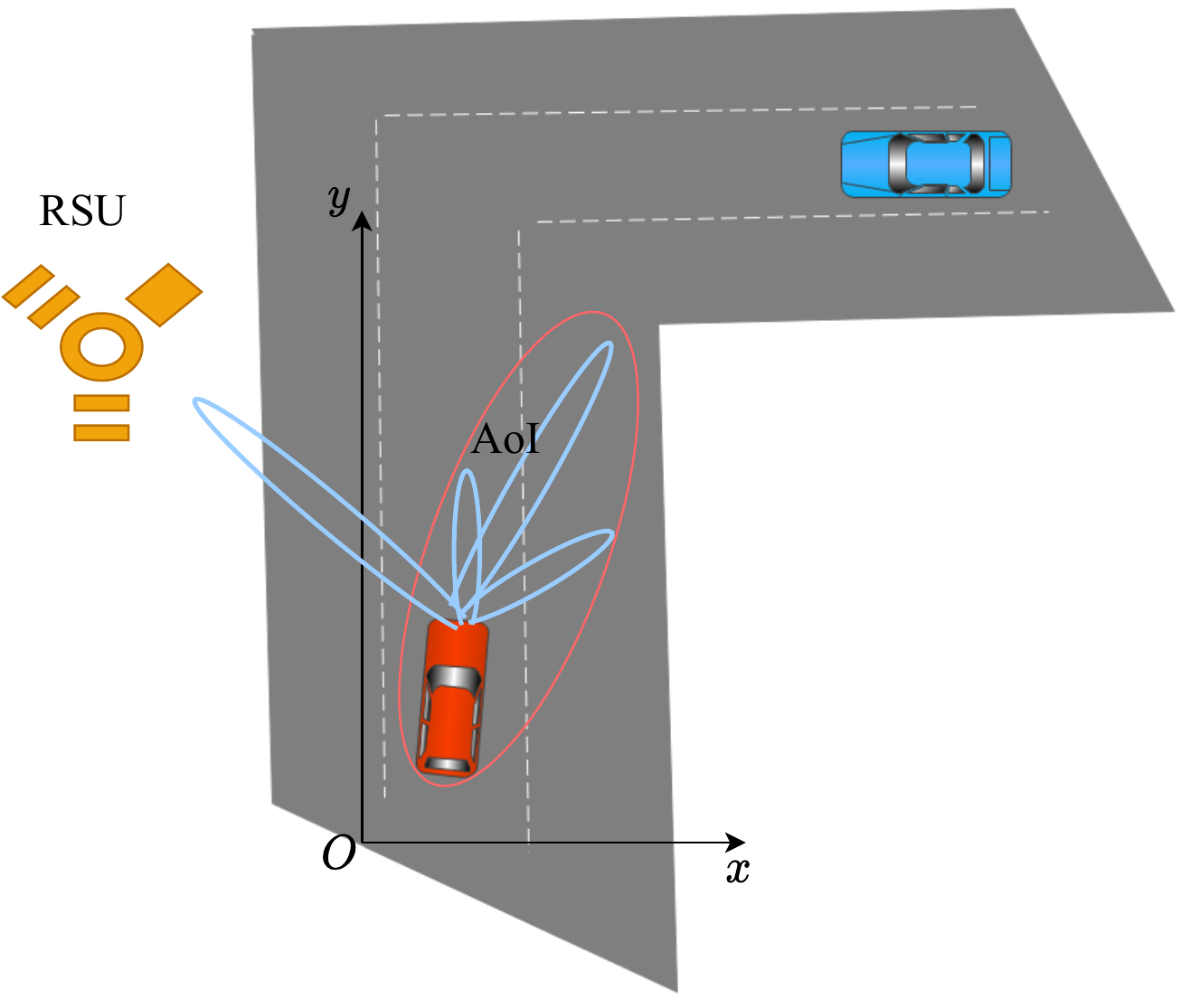}
  \caption{Application scenario: a vehicle-mounted transmitter communicates with an RSU while sensing the AoI for driving safety enhancement.} 
  \label{fig1}
\end{figure}

In this section, we present the system model of VBA-ISAC by expatiating on the sensing model and communication model respectively. Specifically, we first give the kinematic vehicle model based on the vehicle behaviors and predict the AoI. Then, the sensing model is presented by specifying the relation between beampattern design and AoI. Lastly, we adopt the communication model from the perspective of maximizing the spectral efficiency of the transceiver. 
In particular, we consider a scenario where a vehicle-mounted transmitter communicates with an RSU while sensing AoI for driving safety purposes, as depicted in Fig.~\ref{fig1}. In the scenario, the RSU serves as the infrastructure for vehicle-to-infrastructure (V2I) communications. 
In the depicted figure, the red car is about to turn right. Considering the specific vehicular behaviors, it needs to sense the road conditions on the right in advance.

\subsection{Vehicle Kinematic Model}
\begin{figure}
  \centering
  \includegraphics[width=0.3\textwidth]{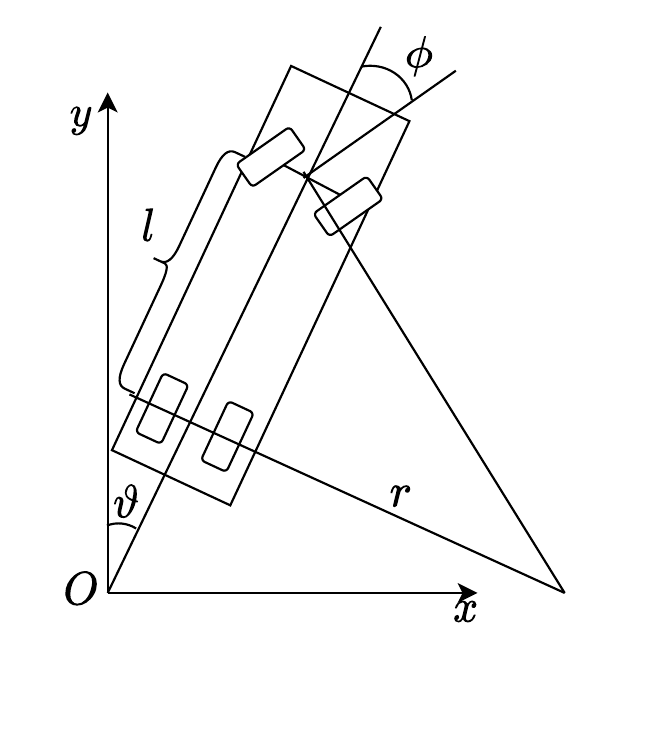}
  \caption{Kinematic model of vehicle referring to the bicycle model.} 
  \label{fig2}
\end{figure}

First, we establish the kinematic model of the vehicle. As shown in Fig.~\ref{fig1}, the red vehicle is driven on the road. At the intersection, it plans to turn right. As the vehicle's height does not affect the following modeling and analysis and is thereby negligible, it can be simplified that the movement of the vehicle is on the x-y (horizontal) plane. For simplicity, we further suppose that the left and right wheels of the vehicle have the same steering angle and velocity at any time \cite{jason2015kinematic}. Overall, for ease of analysis, we model the movement of the vehicle via a bicycle model as demonstrated in Fig.~\ref{fig2}.

Through in-device sensors, the current state of the vehicle can be obtained. Without loss of generality, we denote the distance between the front and rear wheels, displacements on the x-axis and y-axis, velocity, acceleration, driving direction, and steering angle of the vehicle by ${l}$, ${d}^{x}$, ${d}^{y}$, ${v}$, ${a}$, ${\vartheta }$, and ${\phi}$, respectively. The motion state of the vehicle can be expressed as
\begin{equation} 
\mathbf{s}=[{d}^{x},{d}^{y},{v},{\vartheta }]^T.
\end{equation}
In the model, acceleration ${a}$ and steering angle ${\phi}$ are the driver's controllable input. We consider the influence of mechanical inertia, and therefore, ${a}$ and ${\phi}$ remain unchanged in an instant. As a consequence, the movement of the vehicle in an instant is approximate to a circular curve \cite{sun2017fuzzy}. According to the geometric relation depicted in Fig.~\ref{fig2}, the radius can be calculated as
\begin{equation} \label{r}
{r} = \frac{l}{\tan \phi}.
\end{equation}

Furthermore, the vehicle is assumed to move in the direction of the body in an instant \cite{philip2017kinematic}. The moving distance of the vehicle within duration ${\Delta t}$ can be determined as 
\begin{equation} \label{s}
{\Delta s} = {r}{\Delta \vartheta }.
\end{equation}
According to (\ref{r}) and (\ref{s}), the change of driving direction can be calculated as
\begin{equation} \label{angle}
{\Delta \vartheta } =\frac{\tan \phi}{l}{\Delta s}. 
\end{equation}
We divide both sides of (\ref{angle}) by ${\Delta t}$. The rate of change of driving direction can be expressed as
\begin{equation} \label{w}
{\dot \vartheta } =\frac{\tan \phi}{l}{v}.
\end{equation}
In addition, it needs to consider the displacements of the vehicle on the x-axis and y-axis in an instant. As shown in Fig.~\ref{fig2}, it can be obtained as
\begin{equation} \label{dv}
\frac{\Delta {d}^{x}}{\Delta {d}^{y}} ={\tan \vartheta }.
\end{equation}
The velocities on the x-axis and y-axis can be denoted by ${\dot {d}^{x}}=\frac{\Delta {d}^{x}}{\Delta {t}}$ and ${\dot {d}^{y}}=\frac{\Delta {d}^{y}}{\Delta {t}}$. According to (\ref{dv}), the relation between them can be established as
\begin{equation} \label{}
-{\dot {d}^{x}}{\cos \vartheta } + {\dot {d}^{y}}{\sin \vartheta } = 0.
\end{equation}
At this point, we attain a simplified  kinematic vehicle model, summarized as follows
\begin{equation} \label{motion}
\begin{bmatrix}
{\dot \vartheta }\\
{\dot {d}^{x}}\\ 
{\dot {d}^{y}} 
\end{bmatrix} =  \begin{bmatrix}
v\frac{\tan \phi}{l}\\
v\sin \vartheta  \\ 
v\cos \vartheta  
\end{bmatrix}.
\end{equation}
Based on this simplified kinematic vehicle model, given the controllable input ${a}$ and ${\phi}$ at a certain moment, the state information of the vehicle at the next moment can be estimated.

\subsection{Sensing Model}

As shown in Fig.~\ref{fig1} and Fig.~\ref{fig3}, the red vehicle needs to generate beams to sense the AoI for driving safety enhancement. Unlike an omnidirectional MIMO radar, the transmitted beam designed in this paper is allowed to be directional. The transmitter needs to point some beams to the right to sense the AoI. 
For MIMO radar probing purposes, it is desirable to focus the transmit energy on the spatial sections of interest. Hence, the radar beamformer should be designed with good beampattern behavior. 
The radar beampattern located at $\theta$ direction can be written as \cite{cheng2021hybrid}
\begin{equation} \label{beam pattern}
 P(\theta ) = {\mathbf a}_t^H(\theta ){\mathbf{R}_d}{{\mathbf a}_t}(\theta ),
\end{equation}
where ${{\mathbf a}_t}(\theta ) \in {\mathbb{C}^{{N_t} \times 1}}$ is the transmit array response vector; ${N_t}$ is the number of transmit antennas; ${\mathbf{R}_d \in \mathbb{C}^{N_t \times N_t}}$  is the covariance matrix of the radar beampattern. 

In this paper, we consider a uniform linear array (ULA) at the transmitter and the receiver. The array response vector corresponding to ${\theta}$ can be expressed as 
\begin{equation}
{\mathbf{a}}(\theta ) = {\left[ {1,\;{e^{j\frac{2\pi}{\lambda} {d}\sin \theta }},\; \cdots ,\;{e^{j\frac{2\pi}{\lambda} {d}(N - 1)\sin \theta }}}\right]^T},
\end{equation}
where ${N}$ denotes the number of antennas; ${\lambda}$ stands for the wavelength; and ${d}$ stands for the antenna spacing.

The covariance matrix of the radar beampattern can be obtained by the radar beamformer matrix. Specifically, covariance matrix ${\mathbf{R}_d \in \mathbb{C}^{N_t \times N_t}}$ can be expressed as \cite{liu2019hybrid} 
\begin{equation} \label{covariance matrix}
{\mathbf{R}_d} = {\mathbf{F}_{rad}}\mathbf{F}_{rad}^H.
\end{equation}
The radar beamformer matrix ${\mathbf{F}_{rad}} \in\mathbb{C}^{N_t\times K}$ of non-overlapping subarrays can be expressed as \cite{Wilcox2012subarray}
\begin{equation} \label{frad}
{\mathbf{F}_{rad}} = \left [ {\begin{array}{*{20}{l}} {{{\mathbf{v}}_1}}&0& \cdots &0 \\ 0&{{{\mathbf{v}}_2}}&{}&0 \\ \vdots &{}& \ddots & \vdots \\ 0&0& \cdots &{{{\mathbf{v}}_{K}}} \end{array}} \right] \in\mathbb{C}^{N_t\times K},
\end{equation}
where ${{\mathbf{v}}_k} \in {\mathbb{C}^{{N_k} \times 1}}$ $(1\leq k\leq K )$ is the $k$-th sub-array steering vector and can be expressed as
\begin{equation}
{\mathbf{v}}_k = {\left[ {1,\;{e^{j\frac{2\pi}{\lambda} {d}\sin \theta_k }},\; \cdots ,\;{e^{j\frac{2\pi}{\lambda} {d}(N_k - 1)\sin \theta_k }}}\right]^T},
\end{equation}
where ${N_k}$ denotes the number of antennas at the $k$-th pointing angle, and ${\sum_{k=1}^K} {N_k}={N_t}$;
and ${K}$ is the number of radar pointing angles. The radar pointing angle ${\theta}_k$ is determined by the AoI. ${N_k}$ is related to the required sensing distance. In short, the more antennas forming narrow beams, the greater the beam power toward the interesting pointing angle will be. The radar beamformer is adapted by adjusting ${N_k}$ and ${\theta}_k$.

\subsection{Communication Model}
\begin{figure}
  \centering
  \includegraphics[width=0.48\textwidth]{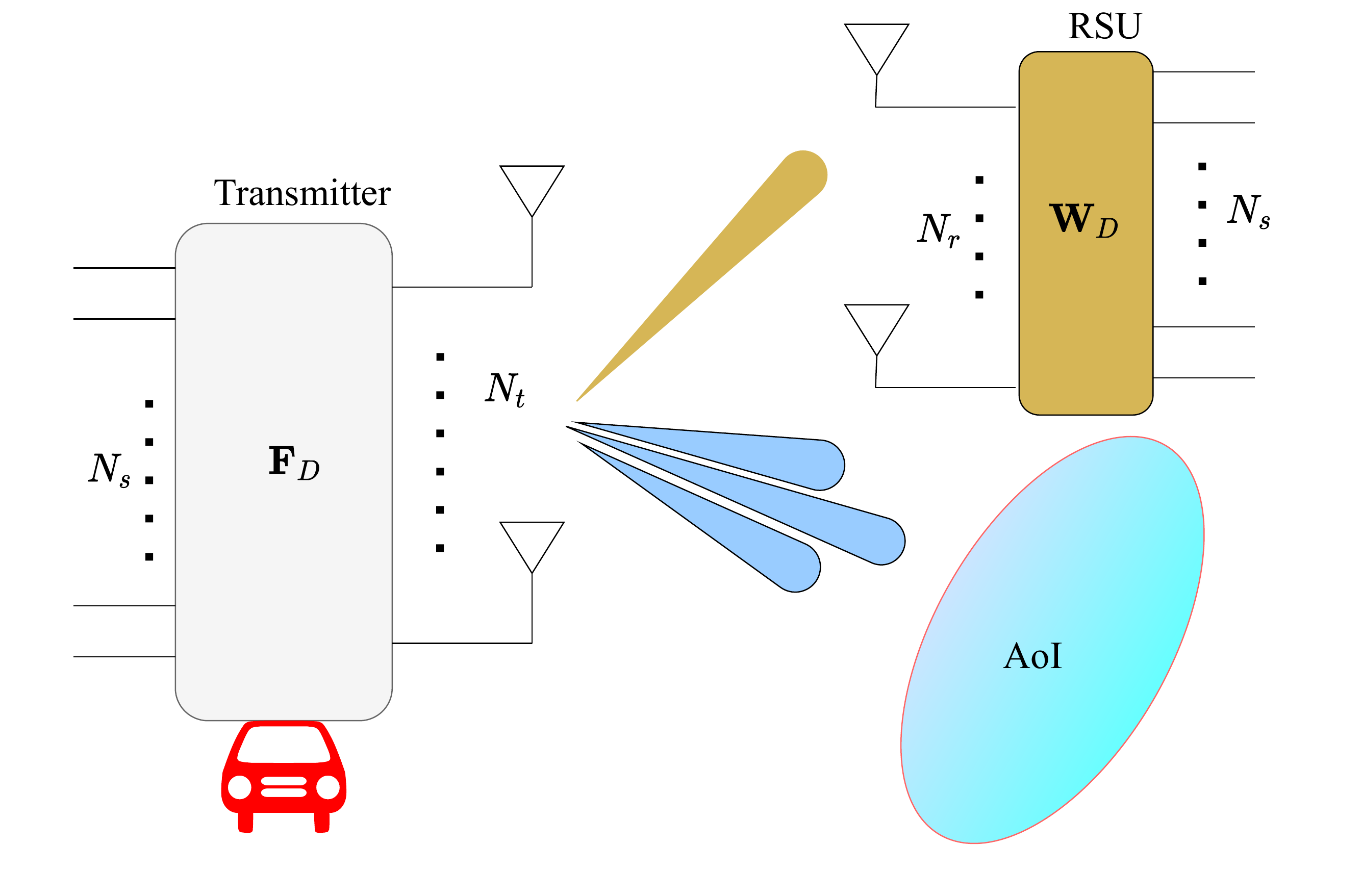}
  \caption{Multiple data streams are transmitted from a vehicle-mounted multi-antenna transmitter to an RSU for simultaneous sensing and communications.} 
  \label{fig3}
\end{figure}

In Fig.~\ref{fig1}, the vehicle-mounted transmitter provides spectral-efficient uplink data services with the help of the RSU. Without loss of generality, it is assumed that the vehicle-mounted transmitter is equipped with $N_t$ transmit antennas, and the RSU is with $N_r$ receive antennas. As shown in Fig.~\ref{fig3}, $N_s$ data streams are transmitted from the vehicle-mounted transmitter to the RSU.  Conditioned on the assumption that both the RSU and the vehicle are equipped with full-digital RF chain structures, the $N_s$ data streams have to pass through a transmit beamformer ${\mathbf{F}_D}\in\mathbb{C}^{N_t\times N_s}$ before being sent by $N_t$ transmit antennas. Through the wireless channels denoted by ${\mathbf{H}}\in\mathbb{C}^{N_r\times N_t}$, the sent radio waves reach the receiving side and then pass through the receive combiner ${{\mathbf{W}}_{D}}\in\mathbb{C}^{N_s\times N_r}$. Let $\mathbf s\in{\mathbb{C}^{N_s\times 1}}$ represent the data symbol vector, and $\mathbb{E}\left[\mathbf{ss}^H\right]=\mathbf{I}_{N_s}$. The normalized power constraint can be expressed as $||{\mathbf{F}_D}||_F^2 = {N_s}$. With the above formulations, the transmitted signal ${\mathbf{x}}$ can be expressed as ${\mathbf{x}} = {\mathbf{F}_D}{\mathbf{s}}$, and the signal on the receiving side can be expressed as 
\begin{equation} \label{y0}
{\mathbf y} = \sqrt{ p} {{\mathbf{W}}_{D}}{\mathbf{H}}{\mathbf{F}_D}{\mathbf{s}} + {{\mathbf{W}}_{D}}{\mathbf{n}},
\end{equation}
where $p$ stands for the average power of the received signal, and $\mathbf{n}\in\mathbb{C}^{N_r\times 1}$  represents the vector of complex additive white Gaussian noises distributed over each element obeying $\mathcal{CN}(0,\sigma_n^2)$.

To meet the high-rate requirements for modern vehicular networks,  the mmWave band is adopted for V2I communications. The high-frequency band of mmWave also enables high-resolution sensing performance. In this paper, the Saleh-Valenzuela model \cite{Saleh1987} is adopted to characterize mmWave channel matrix $\mathbf{H}$ as
\begin{equation} \label{h}
{\mathbf{H}} =  \sum\limits_{l = 1}^L {{\alpha _l}} {{\mathbf{a}}_r}({\theta _{r,l}}){\mathbf{a}}_t^H({\theta _{t,l}}),
\end{equation}
where ${L}$ represents the number of paths through a wireless channel; ${\alpha _l}$ stands for the gain of each path; ${{\mathbf{a}}_r}({\theta _{r,l}})$ and ${\mathbf{a}}_t({\theta _{t,l}})$ denote the array response vectors at the receiving side and the transmitting side, respectively; ${\theta _{r,l}}$ and ${\theta _{t,l}}$ represent the angle of arrival and angle of departure, respectively. In this paper, we assume that the CSI for communications has been fully acquired by channel estimation, which is a commonly accepted assumption in the literature of mmWave and vehicular communications \cite{liu2018toward,liu2019hybrid,dong2022vpaitmin}.

Considering that the transmitter and the receiver of vehicular communication systems are distant compared to the wavelength, it is reasonable to assume that the receiving side is with an optimal combiner, denoted by ${{\mathbf{W}}_{opt}}$. By replacing $\mathbf{W}_{D}$ with ${{\mathbf{W}}_{opt}}$ in (\ref{y0}), the received signal can be rewritten as
\begin{equation}  \label{y}
{\mathbf y} = \sqrt{ p} {{\mathbf{W}}_{opt}}{\mathbf{H}}{\mathbf{F}_D}{\mathbf{s}} + {{\mathbf{W}}_{opt}}{\mathbf{n}}.
\end{equation}

\section{VBA-ISAC Beamforming Designs for Vehicle-Mounted Multi-Antenna Transmitter} \label{III}
For the application scenario shown in Fig.~\ref{fig1}, we have formulated the vehicle's kinematic model, sensing model, and communication model in the last section, based on which we propose the VBA-ISAC beamforming designs for vehicle-mounted multi-antenna transmitter in this section.

\subsection{AoI Prediction}
Based on the kinematic model and the current vehicular state, the state of the vehicle in the next instant can be predicted. The displacement on the x-axis in the next instant can be expressed as 
\begin{equation} \label{dx}
{d}^{x} = \int_{t_0}^{{t_0+\Delta t}} {\dot {d}^{x}} \text{d}t = \int_{t_0}^{{t_0+\Delta t}} \sin (\vartheta _{t_0} + {\dot \vartheta }t )({v}_{t_0}+{a}{t})\text{d}t,
\end{equation}
where ${t_0}$ represents the initial instant; $\vartheta _{t_0}$ represents driving direction at ${t_0}$; ${v}_{t_0}$ represents velocity at ${t_0}$. In the same way, the displacement on the y-axis can be expressed, ditto, as
\begin{equation} \label{dy}
{d}^{y} = \int_{t_0}^{{t_0+\Delta t}} {\dot {d}^{y}} \text{d}t = \int_{t_0}^{{t_0+\Delta t}} \cos (\vartheta _{t_0} + {\dot \vartheta }t )({v}_{t_0}+{a}{t})\text{d}t.
\end{equation}
According to the above derivations, we can forecast the position of the vehicle in the next instant. As a continuous driving process, the driving path of the vehicle within a short period of time is thus predictable. The driving path of the vehicle on the two-dimensional plane can be expressed as
\begin{equation} \label{fy}
{d}^{y} = f_c({d}^{x}),
\end{equation}
where $f_c(\cdot)$ is the curve function on the two-dimensional axis. To avoid vehicle collisions, we assume a vehicle safety zone, which is a circle with a radius ${r_s}$. Consequently, the AoI needed to be predicted can be simplified as the area covered by the circle moving, and the trajectory of the circular center is curvilinear (\ref{fy}). 

\begin{figure}
  \centering
  \includegraphics[width=0.34\textwidth]{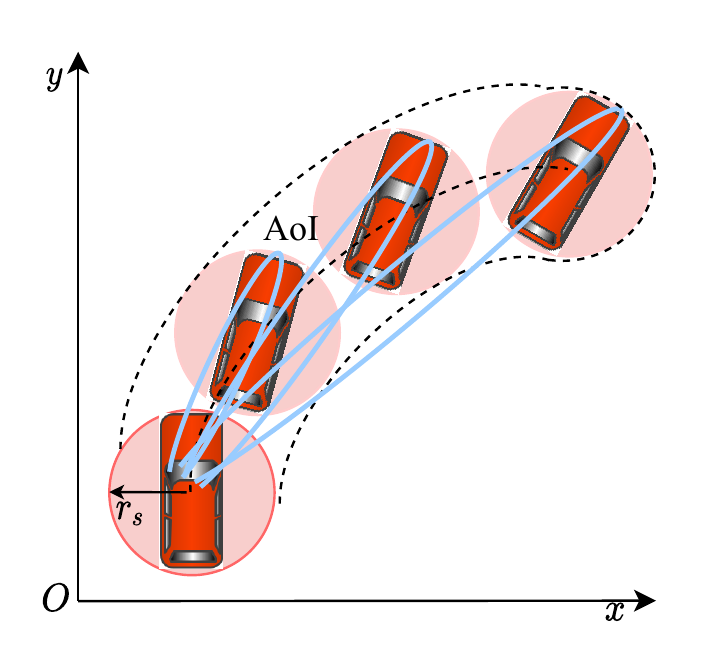}
  \caption{An illustration of the AoI prediction based on the real-time behaviors and the state of the vehicle.} 
  \label{fig4}
\end{figure}

According to the above description, we can predict the AoI that needs to be sensed according to the real-time behaviors and the state of the vehicle. For clarity, the AoI is illustrated in Fig.~\ref{fig4}.

\subsection{Desired Radar Beamformer Calculation}

From (\ref{frad}), it is known that the desired radar beamformer is determined by ${\theta}_k$ and ${N_k}$. To get these parameters, we evenly divide the driving process of the vehicle within ${\Delta t}$ into $K$ stages so that there are $K$ vehicular positions in AoI.
The $K$ beams are used to sense these $K$ positions to fully cover the AoI, as shown in Fig.~\ref{fig4}. The pointing angle ${\theta}_k$ of $k$-th sub-array steering vector ${{\mathbf{v}}_k}$ can be calculated as
\begin{equation} \label{theta}
{\theta}_k = \arctan \left(\frac{{d}^{x}_{k}}{{d}^{y}_{k}} \right ),
\end{equation}
where ${d}^{y}_{k}$ and ${d}^{x}_{k}$ stand for the distances on the x-axis and y-axis of the $k$-th position relative to the initial position of the vehicle, respectively.

 For the MIMO radar sensing, the effective sensing range is positively correlated with the peak value of the main lobe of the beam. The main lobe of the beam can be made narrower by adjusting the number of antennas to achieve a longer radar sensing range. According to the MIMO radar equation \cite{li2011radarequation}, the maximum  radar range can be expressed as
\begin{equation} 
{{d}^{max}_{k}} = \left ( \frac{{P}_{k}{G}^2{\lambda}^2{S}_{\sigma}}{{(4 \pi)}^3{P}_{min}} \right )^{\frac{1}{4}} = {\Omega} \left ( {P}_{k} \right )^{\frac{1}{4}},
\end{equation}
where ${P}_{k}$ represents the radar transmit power at the $k$-th beam, which is determined by the radar beamformer; ${G}$ represents antenna gain; ${\lambda}$ stands for wavelength; ${S}_{\sigma}$ stands for radar cross section; ${P}_{min}$ stands for the minimum detectable signal power, and we denote ${\Omega }=  \left ( \frac{{G}^2{\lambda}^2{S}_{\sigma}}{{(4 \pi)}^3{P}_{min}} \right )^{\frac{1}{4}}$ for simplicity. 

Assuming the total power is evenly distributed to all antennas and the transmit power of each antenna is ${P}_0$, the radar transmit power at the $k$-th beam can be expressed as \cite{li2011radarequation} 
\begin{equation} 
{P}_{k} ={{N_k}{P}_0}.
\end{equation}

The detection distance at the pointing angle of interest ${\theta}_k$ required by the vehicle is determined by the AoI. According to the vehicle's kinematic model, the sensing distance at the pointing angle of interest ${\theta}_k$ can be expressed as
\begin{equation} \label{dsk}
{d}^{s}_k = \sqrt{({d}^{x}_k)^2+({d}^{y}_k)^2}+{r_s}.
\end{equation}
For safe and autonomous driving, it should make ${d}^{max}_k \geq  {d}^{s}_k$ by adjusting ${N_k}$.

 Through the above analysis, we have established the link between the radar beampatterns and the behavior of the vehicle. Once an AoI is selected, the desired beamformer ${\mathbf{F}_{rad}}$ can be obtained. To yield the desired beampattern for a given AoI, the transmit beamformer ${\mathbf{F}_D}$ needs to be designated to approach ${\mathbf{F}_{rad}}$. To simplify the formulated optimization problem, we define that the dimensions of ${\mathbf{F}_D}$ are equal to those of ${\mathbf{F}_{rad}}$, which can be easily satisfied by multiplying ${\mathbf{F}_{rad}}$ with a unitary matrix \cite{liu2019hybrid}. Mathematically, the optimization can be modeled as
 \begin{equation}
 ||\mathbf{F}_D- \mathbf{F}_{rad}||_F^2 \leq \varepsilon_r,
 \end{equation}
where ${\varepsilon}_r $ is a threshold parameter to control the level of similarity between ${\mathbf{F}_D}$ and ${\mathbf{F}_{rad}}$. 
After the above analysis, we have formulated the joint sensing and beamformer design problem. 
The covariance matrix of the transmitted beamforming can be expressed as
\begin{equation}
\begin{split}
{\mathbf{R}_d}
&= {\mathbb{E}}({\mathbf{F}_D}{\mathbf{ss}^H}{\mathbf{F}}_{D}^H) =  {\mathbf{F}_D}{\mathbb{E}}({\mathbf{s}}{{\mathbf{s}}^H}){\mathbf{F}}_{D}^H \\ 
&= {\mathbf{F}_D}{\mathbf{F}}_{D}^H,
\end{split}
\end{equation}
which determines the power distributed in the space.

\subsection{Optimal Transmit Beamforming Calculation}
The design of ${\mathbf{F}_D}$ also directly affects communication performance. In this work, we utilize spectral efficiency as the communication performance metric. The spectral efficiency of the above system model can be formulated as \cite{Archit2018mod}
\begin{equation}\label{R}
\begin{split}
R=\log\left(\det\bigg(\mathbf{I}_{N_s}+\frac{p}{\sigma_n^2 }{{{{{\mathbf{W}}_{opt}}} }{\mathbf{H}}{\mathbf{F}_D}}{{{\mathbf{F}}^H_{D}}{{\mathbf{H}}^H} {{{\mathbf{W}}_{opt}^H}} }\bigg)\right),
\end{split}
\end{equation}
and is upper bounded by
\begin{equation}\label{R optimal}
\begin{split}
{R_{up}}=\log\left(\det\bigg(\mathbf{I}_{N_s}+\frac{p}{\sigma_n^2 }{{{ {{\mathbf{W}}_{opt}} } }{\mathbf{H}}{{\mathbf{F}}_{opt}}}
{{{\mathbf{F}}^H_{opt}}{{\mathbf{H}}^H} {{{\mathbf{W}}_{opt}^H}} }\bigg)\right),
\end{split}
\end{equation}
where  ${\mathbf{F}_{opt}}$ represents the optimal beamforming matrix. ${\mathbf{W}_{opt}}$ and ${\mathbf{F}_{opt}}$ can be obtained by performing singular value decomposition on channel matrix ${\mathbf{H}}$ \cite{yu2016alter}.

To maximize the spectral efficiency of VBA-ISAC systems, the beamformer can be designed to minimize the Euclidean distance between ${\mathbf{F}_D}$ and ${\mathbf{F}_{opt}}$. We formulate it as
\begin{equation} \label{fopt}
||{\mathbf{F}_D}- {{\mathbf{F}}_{opt}}||_F^2 \leq {\varepsilon}_c, 
\end{equation}
where ${\varepsilon}_c$ is a threshold parameter to control the level of the similarity between ${\mathbf{F}_D}$ and ${\mathbf{F}_{opt}}$.

\subsection{Problem Formulation for VBA-ISAC}

According to the above communication model and sensing model, to optimize the VBA-ISAC system, we need to minimize the Euclidean distances between ${\mathbf{F}_D}$ and ${{\mathbf{F}}_{opt}}$ as well as ${\mathbf{F}_{rad}}$ at the same time. However, the optimization of these two distances is not compatible. Such incompatibility results from the performance trade-off between communications and sensing, which is worth considering and devising. In this paper, we strike this trade-off by introducing a trade-off factor $\rho$, which is ranging from $0$ to $1$. The joint communication and sensing optimization problem of beamformer design can be formulated as
\begin{equation}\label{model1}
\begin{array}{l}
\mathop {\min }\limits_{_{{\mathbf{F}_{D}}}} \;\rho||{\mathbf{F}_D} - {{\mathbf{F}}_{opt}}||_F^2 + (1 - \rho)||{{\mathbf F}_{D}} - {\mathbf{F}_{rad}}||_F^2 \\  
\;s.t. \;||{\mathbf{F}_D}||_F^2 = {N_s}.
\end{array}\end{equation}

From the form of (\ref{model1}), it is obviously a quadratic constraint quadratic programming (QCQP) optimization problem. It is non-convex and intricate to be optimally solved. Fortunately, the classic semi-definite relaxation (SDR) algorithm, as a commonly used tool in the fields of communications and signal processing, can be leveraged to obtain approximate and sub-optimal solutions to QCQP optimization problems \cite{liu2018toward}.

To apply the SDR algorithm to solve the joint optimization problem formulated in (\ref{model1}), we first simplify (\ref{model1}) to make it look more concise and facilitate subsequent analysis. It can be converted into the following form
\begin{equation}\begin{array}{l} \label{Fab}
\mathop {\min }\limits_{_{{\mathbf{F}_D}{}}} ||{{\mathbf{A}}}{\mathbf{F}_D} - {{\mathbf{B}}}||_F^2 \\  
\;s.t. \;||{\mathbf{F}_D}||_F^2 = {N_s},
\end{array} 
\end{equation}
where ${\mathbf{A}} = {[ {\sqrt \rho  {{\mathbf{I}}^T_{N_t}},\sqrt {1 - \rho } {\mathbf{I}}^T_{N_t}} ]^T}$, and ${\mathbf{B}} = {[ {\sqrt \rho  {{\mathbf{F}}^T_{opt}},\sqrt {1 - \rho } \mathbf{F}_{rad}^T}]^T}$.
Based on the proven theorems and transformations given in \cite{luo2010semidef}, we reformulate (\ref{Fab}) as a homogeneous QCQP optimization problem
\begin{equation}\label{qcqp}
\begin{array}{l}
\mathop {\min }\limits_{_{{\mathbf{X}}{}}} \mathrm{Tr}(\mathbf{CX}) \\ 
s.t. \; \mathrm{Tr}(\mathbf{A}_1\mathbf{X})={N_s}\\
\;\;\;\;\;\; \mathrm{Tr}(\mathbf{A}_2\mathbf{X})=1\\
\;\;\;\;\;\; \mathbf{X}\succeq 0,\;\mathrm{rank}(\mathbf{X})=1,
\end{array}
\end{equation}
where $\mathbf{X}=\mathbf{f}_{D}{\mathbf{f}^H_{D}}$ is an $({N_t}{N_s}+1)$-dimension complex Hermitian matrix with $\mathbf{f}_{D}=[{{\mathrm{vec}(\mathbf{F}_{D})}\;\;{t}}]^T$ in which ${t}$ is used to judge whether the optimal solution is $\mathbf{f}_{D}$ or $-\mathbf{f}_{D}$, i.e., $t = 1$ or $t = -1$; ${\mathbf{A}_1}$, ${\mathbf{A}_2}$, and ${\mathbf{C}}$ are given by
\begin{equation*}
\mathbf{A}_1=\left[ {\begin{array}{*{20}{c}}
{\mathbf{I}_{{N_t}{N_s}}}&{\mathbf{0}}\\
{\mathbf{0}}&{0}
\end{array}} \right],
\mathbf{A}_2=\left[ {\begin{array}{*{20}{c}}
{\mathbf{0}_{{N_t}{N_s}}}&{\mathbf{0}}\\
{\mathbf{0}}&{1}
\end{array}} \right],
\end{equation*}
and
\begin{equation*}
\mathbf{C}=\left[ {\begin{array}{*{20}{c}}
{(\mathbf{I}_{N_s}\otimes\mathbf{A})^H(\mathbf{I}_{N_s}\otimes\mathbf{A})}&
{-(\mathbf{I}_{N_s}\otimes\mathbf{A})^H{\mathrm{vec}(\mathbf{B})}}\\
{-{\mathrm{vec}(\mathbf{B})}^H(\mathbf{I}_{N_s}\otimes\mathbf{A})}&
{{\mathrm{vec}(\mathbf{B})}^H{\mathrm{vec}(\mathbf{B})}}
\end{array}} \right],
\end{equation*}
where the dimensions are both $\mathbb{C}^{({N_t}{N_s}+1)\times ({N_t}{N_s}+1)}$.

However, due to the rank-one constraint, (\ref{qcqp}) is still non-convex. After removing the rank-one constraint, it becomes a semi-definite programming (SDP) problem
\begin{equation}\label{sdp}
\begin{array}{l}
\mathop {\min }\limits_{_{{\mathbf{X}}{}}} \mathrm{Tr}(\mathbf{CX}) \\ 
s.t. \; \mathrm{Tr}(\mathbf{A}_1\mathbf{X})={N_s}\\
\;\;\;\;\;\; \mathrm{Tr}(\mathbf{A}_2\mathbf{X})=1\\
\;\;\;\;\;\; \mathbf{X}\succeq 0.
\end{array}
\end{equation}
(\ref{sdp}) becomes a classic convex optimization problem, which can be solved by conventional convex optimization tools, such as CVX toolbox. In (\ref{sdp}), we relax the rank-one constraint, resulting in the globally optimal solution to (\ref{sdp}). However, the global optimum of (\ref{sdp}) is not equivalent to an even feasible solution to (\ref{Fab}) when the rank of $\mathbf{X}$ is higher than one. Utilizing the approach proposed in \cite{luo2010semidef}, we apply the eigen-decomposition of $\mathbf{X}$ to yield  an approximate solution to (\ref{Fab}). We summarize the approach tailored for our formulated optimization problem as in Algorithm 1.

\renewcommand{\algorithmicrequire}{\textbf{Input:}}
\renewcommand{\algorithmicensure}{\textbf{Output:}}
\begin{algorithm}[t] 
\caption{Semi-definite relaxation (SDR) algorithm for VBA-ISAC beamforming design.}
\label{alg:B}
\begin{algorithmic}
    \REQUIRE ${{\mathbf{F}}_{opt}}, {\mathbf{F}_{rad}}, {N_s}, {\varepsilon}_{r,c} >0, 0 \leqslant \rho \leqslant 1$
    
    \STATE Randomly initialize ${\mathbf{F}_D}$.
    \STATE 1. Convert (\ref{model1}) into (\ref{Fab}). Then reformulate it as a homogeneous QCQP optimization problem as (\ref{qcqp}).
    \STATE 2. Relax the rank-one constraint in (\ref{qcqp}) and turn it into an SDP problem as (\ref{sdp}). 
    \STATE 3. Solve the SDP problem by using convex optimization toolbox CVX, 
    \STATE 4. Apply the eigen-decomposition of $\mathbf{X}$ to yield an approximate and suboptimal solution to the original optimization problem.

\ENSURE ${\mathbf{F}_D}$ 
\end{algorithmic}
\end{algorithm}

\subsection{VBA-ISAC Algorithm and Computational Complexity Analysis}
Through the above analysis, we can design the ISAC beamforming based on the vehicle's behavior and state. We summarize the VBA-ISAC beamforming scheme as in Algorithm 2.

\renewcommand{\algorithmicrequire}{\textbf{Input:}}
\renewcommand{\algorithmicensure}{\textbf{Output:}}
\begin{algorithm} [t] 
\caption{VBA-ISAC beamforming scheme.}
\label{alg:B}
\begin{algorithmic}
    \REQUIRE Acceleration ${a}$ and steering angle ${\phi}$
    \STATE 1. Predict the AoI according to the real-time behavior and state of the vehicle from in-vehicle sensors.
    \STATE 2. Formulate the desired radar beamformer ${\mathbf{F}_{rad}}$ based on the predicted AoI.
    \STATE 3. Obtain the optimal transmit beamformer ${\mathbf{F}_{opt}}$ by performing singular value decomposition on channel matrix ${\mathbf{H}}$.
    \STATE 4. Formulate the joint optimization problem by introducing a trade-off factor to balance the communication and sensing performance.
    \STATE 5. Solve the Formulate the joint optimization problem by applying the SDR algorithm.
    \ENSURE The desired VBA-ISAC beamforming matrix. 
\end{algorithmic}
\end{algorithm}

According to the proposed scheme, the AoI is obtained by calculating the value of a one-dimensional function. The computational complexity of such calculation can be omitted in general because the complexity of the entire scheme is dominated by the SDR algorithm.  According to the methodology adopted in \cite{luo2010semidef}, the worst-case complexity of solving (\ref{sdp}) is ${\mathcal{O}}\left({N_t}^{3.5}{N_s}^{3.5}\right)$.

\section{VBA-ISAC Beamforming Design with the Hybrid Architecture} \label{IV}

For multi-antennas ISAC systems, full-digital beamforming demands RF chains, including signal mixers and analog-to-digital converters, comparable in number to the antenna elements. The prohibitive cost and power consumption of RF chains make full-digital ISAC beamforming design uneconomical and difficult to apply in the practical vehicular system. To reduce the hardware complexity and the associated costs, the hybrid analog-digital (HAD) beamforming structure is more suitable for vehicle-mounted ISAC systems, which requires much fewer RF chains compared to full-digital transceivers \cite{zhao2021partially}. 
Therefore, it makes perfect sense that HAD beamforming design is an attractive technology for practical vehicular ISAC systems. In this section, we study the VBA-ISAC beamforming design with a HAD architecture.

\subsection{Problem Formulation for VBA-ISAC with a HAD Architecture}

\begin{figure*}
	\centering
	\includegraphics[width=0.7\linewidth]{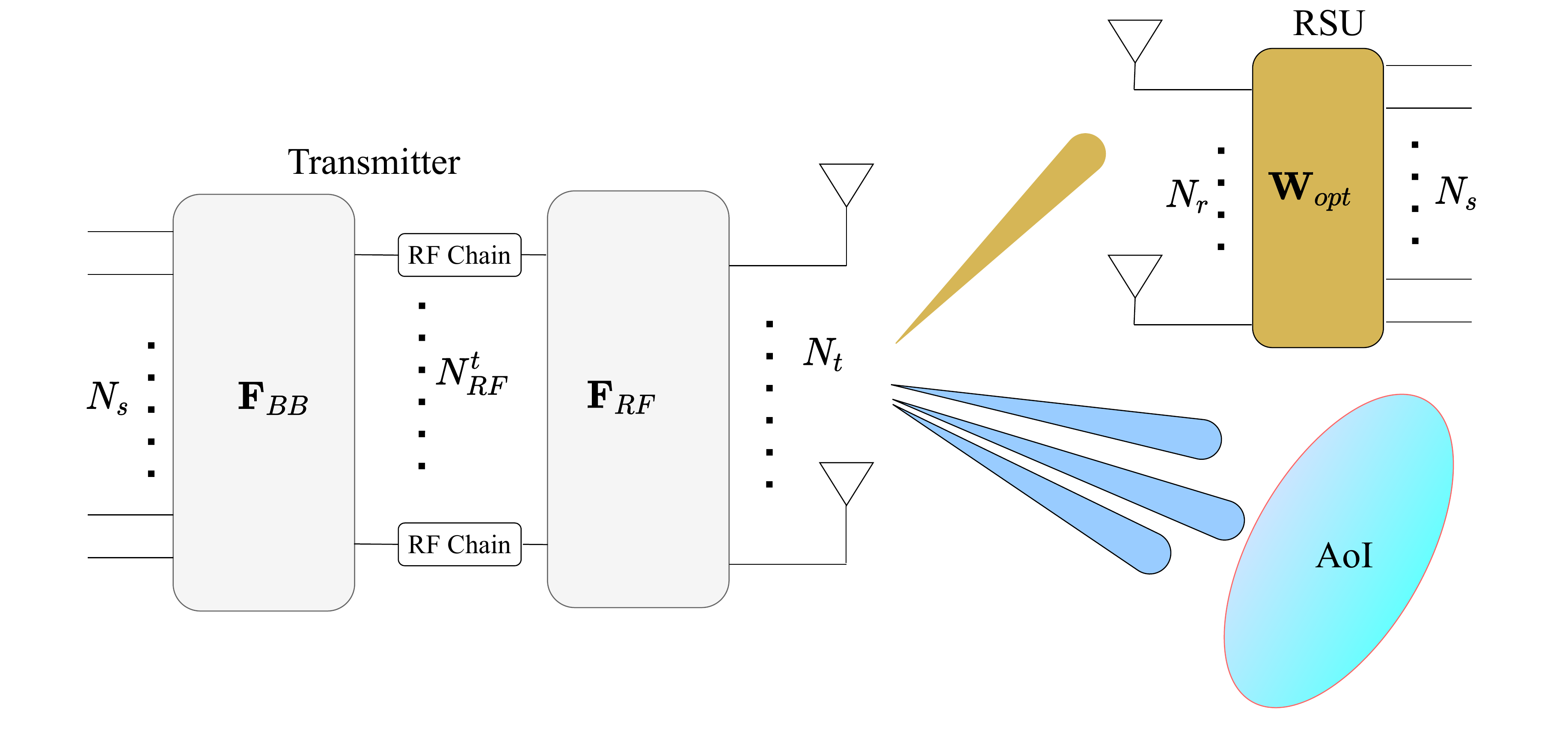}
	\caption{A vehicle-mounted transmitter with HAD architecture communicates with an RSU while sensing the AoI.}
	\label{fig5}	
    \vspace{-0.2cm}
\end{figure*}

Similarly, we present a VBA-ISAC beamforming design scenario with a HAD architecture as depicted in Fig.~\ref{fig5}. The main difference in introducing the HAD architecture is that the transmitting beamformer is now composed of a digital beamformer and an analog beamformer. 
We assume the receiving side with an optimal combiner, denoted by ${{\mathbf{W}}_{opt}}$.

In Fig.~\ref{fig5}, the vehicle-mounted transmitter is equipped with ${N}^{t}_{RF}$ RF chains. Each RF chain is connected to all antennas through phase shifters. The number of RF chains is limited as to be ${N_s} \leqslant  {N}^{t}_{RF} \leqslant {N_t}$. 
Accordingly, transmitted signal vector ${\mathbf{x}}$ can now be expressed as ${\mathbf{x}} = {{\mathbf{F}}_{RF}}{{\mathbf{F}}_{BB}}{\mathbf{s}}$. The normalized power constraint can be expressed as $||{{\mathbf{F}}_{RF}}{{\mathbf{F}}_{BB}}||_F^2 = {N_s}$. The signal on the receiving side is thus given by 
\begin{equation}
{\mathbf y} = \sqrt{p} {{\mathbf{W}}_{opt}}{\mathbf{H}}{{\mathbf{F}}_{RF}}{{\mathbf{F}}_{BB}}{\mathbf{s}} + {{\mathbf{W}}_{opt}} {\mathbf n},
\end{equation}
where $\mathbf{F}_{RF} \in {\mathbb{C}^{N_t\times{N}^{t}_{RF}}}$ is the analog beamformer; and $\mathbf{F}_{BB} \in {\mathbb{C}^{{N}^{t}_{RF}\times N_s}}$ is the digital beamformer.

Exactly the same as full-digital beamforming VBA-ISAC systems, the spectral efficiency of the above communication model considering the HAD architecture can be formulated as
\begin{equation}\label{R}
\begin{split}
R=\log\bigg(\det\bigg(\mathbf{I}_{N_s}+\frac{p}{\sigma_n^2 }{{{ {\mathbf{W}}_{opt} } }{\mathbf{H}}{{\mathbf{F}}_{RF}}{{\mathbf{F}}_{BB}}}\\
\times{{\mathbf{F}}_{BB}^H{\mathbf{F}}_{RF}^H{{\mathbf{H}}^H} {{{\mathbf{W}}^H_{opt}}} }\bigg)\bigg).
\end{split}
\end{equation}

Similarly, to maximize the spectral efficiency of hybrid VBA-ISAC systems, we should design the hybrid beamformer to achieve the smallest Euclidean distance between ${{\mathbf{F}}_{RF}}{{\mathbf{F}}_{BB}}$ and ${\mathbf{F}_{opt}}$, which can be explicitly expressed as
\begin{equation}
||{{\mathbf{F}}_{RF}}{{\mathbf{F}}_{BB}} - {{\mathbf{F}}_{opt}}||_F^2 \leq {\varepsilon}_c. 
\end{equation}

For the sensing model with the HAD architecture, it is similar to full-digital beamformer VBA-ISAC systems, and the desired radar beampattern is designed based on AoI. The radar covariance matrix of HAD beamforming can be expressed as
\begin{equation}
\begin{split}
{\mathbf{R}_d}
&= {\mathbb{E}}({{\mathbf{F}}_{RF}}{{\mathbf{F}}_{BB}}{\mathbf{ss}^H}{\mathbf{F}}_{BB}^H{\mathbf{F}}_{RF}^H)\\ 
&=  {{\mathbf{F}}_{RF}}{{\mathbf{F}}_{BB}}{\mathbb{E}}({\mathbf{s}}{{\mathbf{s}}^H}){\mathbf{F}}_{BB}^H{\mathbf{F}}_{RF}^H \\ 
&= {{\mathbf{F}}_{RF}}{{\mathbf{F}}_{BB}}{\mathbf{F}}_{BB}^H{\mathbf{F}}_{RF}^H. 
\end{split}
\end{equation}

Similarly, to generate a satisfying beampattern, we need to design the hybrid beamformer to achieve the smallest Euclidean distance between ${{\mathbf{F}}_{RF}}{{\mathbf{F}}_{BB}}$ and ${\mathbf{F}_{rad}}$, which can be similarly expressed as
\begin{equation}
||{{\mathbf{F}}_{RF}}{{\mathbf{F}}_{BB}} - {\mathbf{F}_{rad}}||_F^2 \leq {\varepsilon}_r.
\end{equation}
For simplicity, again, we assume that the dimensions of ${{\mathbf{F}}_{RF}}{{\mathbf{F}}_{BB}}$ are equal to ${\mathbf{F}_{rad}}$.

According to the above communication model and sensing model with the HAD architecture, the formulation of hybrid beamforming VBA-ISAC systems can be written as
\begin{equation}\label{hybrid model}
\begin{array}{l}
\mathop {\min }\limits_{_{{{\mathbf{F}}_{RF}},{{\mathbf{F}}_{BB}}}} \rho||{{\mathbf{F}}_{RF}}{{\mathbf{F}}_{BB}} - {{\mathbf{F}}_{opt}}||_F^2 \\
\;\;\;\;\;\;\;\;\;\;\;\;\;+(1 - \rho)||{{\mathbf F}_{RF}}{{\mathbf{F}}_{BB}} - {\mathbf{F}_{rad}}||_F^2 \\  
\;\;\;s.t. \; \vert {\mathbf{F}_{RF}}_{i,j}\vert =1, \forall i,j \\
\;\;\;\;\;\;\;\;\; ||{{\mathbf{F}}_{RF}}{{\mathbf{F}}_{BB}}||_F^2 = {N_s}.
\end{array}\end{equation}
Since the phase shifters can only adjust the signal phase, not the signal amplitude, ${{\mathbf{F}}_{RF}}$ has to abide by the unit-modulus constraint.

\subsection{Alternating  Minimization for VBA-ISAC with the Hybrid Architecture}

From the form of (\ref{hybrid model}), the formulated joint optimization problem can be regarded as a matrix factorization problem. Since it has two variables to be optimized, alternating minimization is applicable, which adapts one while fixing the other.

Specifically, when optimizing digital beamformer ${{\mathbf F}_{BB}}$, we should first fix the analog beamformer ${{\mathbf F}_{RF}}$. Therefore, (\ref{hybrid model}) can be rewritten as
\begin{equation}\begin{array}{l} 
\mathop {\min }\limits_{_{{{\mathbf{F}}_{BB}}{}}} \rho||{{\mathbf{F}}_{RF}}{{\mathbf{F}}_{BB}} - {{\mathbf{F}}_{opt}}||_F^2 +(1 - \rho)||{{\mathbf F}_{RF}}{{\mathbf{F}}_{BB}} - {\mathbf{F}_{rad}}||_F^2
\end{array} \end{equation}

To facilitate the following formulation and analysis, it can be further converted to
\begin{equation}\begin{array}{l} \label{fbb}
\mathop {\min }\limits_{_{{{\mathbf{F}}_{BB}}{}}} ||{{\mathbf{A}}}{{\mathbf{F}}_{BB}} - {{\mathbf{B}}}||_F^2 ,
\end{array} \end{equation}
where ${\mathbf{A}} = {[ {\sqrt \rho  {{\mathbf{F}}^T_{RF}},\sqrt {1 - \rho } {\mathbf{F}}^T_{RF}} ]^T}$, and ${\mathbf{B}} = {[ {\sqrt \rho  {{\mathbf{F}}^T_{opt}},\sqrt {1 - \rho } \mathbf{F}_{rad}^T}]^T}$. Now, it becomes obvious that (\ref{fbb}) is a classic matrix factorization problem. Due to the transformation, the problem can be solved by SDR method shown in Section III.D, or least squares (LS) method proposed in \cite{yu2016alter} as
\begin{equation}\label{fab}
\mathbf{F}_{BB}=\mathbf{A}^\dagger\mathbf{B}.
\end{equation}
Regarding the power constraint, we multiply $\frac{\sqrt{N_s}}{\Vert\mathbf{F}_{RF}\mathbf{F}_{BB}\Vert_F}$ by the final optimized result.

Similarly, when optimizing analog beamformer ${{\mathbf F}_{RF}}$, we need to fix digital beamformer ${{\mathbf F}_{BB}}$, and, hence, (\ref{model1}) can be refactored as
\begin{equation}\begin{array}{l}\label{frf}
\mathop {\min }\limits_{_{{{\mathbf{F}}_{RF}}{}}} \rho||{{\mathbf{F}}_{RF}}{{\mathbf{F}}_{BB}} - {{\mathbf{F}}_{opt}}||_F^2 +(1 - \rho)||{{\mathbf F}_{RF}}{{\mathbf{F}}_{BB}} - {\mathbf{F}_{rad}}||_F^2 \\
s.t. \; \vert {\mathbf{F}_{RF}}_{i,j}\vert =1, \forall i,j .
\end{array} \end{equation}

However, (\ref{frf}), as a non-convex optimization problem, is difficult to tackle. Constraint $\vert {\mathbf{F}_{RF}}_{i,j}\vert =1, \forall i,j$ represents the unit-modulus constraint, which cannot be solved by conventional optimization algorithms. Fortunately, it can be transformed into a typical manifold structure, which can subsequently be solved as a manifold optimization problem \cite{yu2016alter,jung2017low,long2017coordinated}. Specifically, we transform (\ref{frf}) to optimization problem on manifold as follows:
\begin{equation}
\begin{split} \label{fx}
\mathop {\min }\limits_{_{{{\mathbf{F}}_{RF}}\in\mathcal{M}}} {f}_m({{\mathbf{F}}_{RF}})= &\rho||{{\mathbf{F}}_{RF}}{{\mathbf{F}}_{BB}} - {{\mathbf{F}}_{opt}}||_F^2 \\
&+(1 - \rho)||{{\mathbf F}_{RF}}{{\mathbf{F}}_{BB}} - {\mathbf{F}_{rad}}||_F^2,
\end{split}
\end{equation}
where ${\mathcal{M}}$ stands for the manifold, and ${f}_m(\cdot)$ is the objective function on manifold. 
Accordingly, inspired by the method utilized in \cite{yu2016alter,jung2017low}, we consider a complex circle manifold of the vector $\mathbf{p} =\mathrm{vec}({{\mathbf{F}}_{RF}})$, which can be expressed as 
\begin{equation} \label{ma}
\mathcal{M}_{cc}=\left\{\mathbf{p}\in \mathbb{C}^{N_t {N}^t_{RF}}: \vert \mathbf{p}_{i}\vert =1, i=1,2, \ldots, {N_t {N}^t_{RF}}\right\}, 
\end{equation}
where $\mathbf{p}$ is a point on the manifold.

According to the defined complex circle manifold given in (\ref{ma}), the optimization problem (\ref{fx}) can be expressed as
\begin{equation}
\begin{split} \label{fxf}
\mathop {\min }\limits_{_{{\mathbf{p}}\in\mathcal{M}_{cc}}} {f}_m(\mathbf{p})= &\rho||{{\mathbf{f}}_{BB}}{\mathbf{p}} - {{\mathbf{f}}_{opt}}||_F^2 \\  
&+(1 - \rho)||{{\mathbf{f}}_{BB}}{\mathbf{p}} - {\mathbf{f}_{rad}}||_F^2 ,
\end{split}
\end{equation}
where ${{\mathbf{f}}_{BB}} = \mathbf{F}^*_{BB} \otimes\mathbf{I}_{N_t}$,  ${{\mathbf{f}}_{opt}} = \mathrm{vec}({{\mathbf{F}}_{opt}})$, and ${\mathbf{f}_{rad}} = \mathrm{vec}({\mathbf{F}_{rad}})$. 
According to \cite{yu2016alter,jung2017low}, (\ref{fxf}) can be solved well by the Manopt toolbox.

\subsection{VBA-ISAC Algorithm with the Hybrid Architecture and Computational Complexity Analysis}

According to the above description, the solution of VBA-ISAC beamforming design with the hybrid architecture is summarized as in Algorithm 3.

\renewcommand{\algorithmicrequire}{\textbf{Input:}}
\renewcommand{\algorithmicensure}{\textbf{Output:}}
\begin{algorithm}[t]
\caption{Alternating minimization algorithm for VBA-ISAC beamforming design with the hybrid architecture.}
\label{alg:B}
\begin{algorithmic}
    \REQUIRE ${{\mathbf{F}}_{opt}}, {\mathbf{F}_{rad}}, {N_s}, \varepsilon_{r,c} >0, 0 \leqslant \rho \leqslant 1, i_{max}>0$

    Randomly initialize ${{\mathbf{F}}^0_{RF}}$ and ${{\mathbf{F}}^0_{BB}}$, and set $i=0$

    \WHILE{$i\le i_{max}$}
    \STATE 1. Fix ${{\mathbf{F}}^i_{RF}}$, and optimize ${{\mathbf{F}}^{i+1}_{BB}}$ by (\ref{fab})
    \STATE 2. Fix ${{\mathbf{F}}^i_{BB}}$, and optimize ${{\mathbf{F}}^{i+1}_{RF}}$ by the manifold optimization method.
    \STATE 3. $i\leftarrow i+1$.
    \STATE 4. Judge whether the convergence condition is satisfied, and break the while loop if yes.
    \ENDWHILE
\ENSURE ${{\mathbf{F}}^i_{RF}}$ and ${{\mathbf{F}}^i_{BB}}$
\end{algorithmic}
\end{algorithm}

In Algorithm 3, the most critical step is to solve (\ref{frf}) for optimizing ${{\mathbf{F}}_{RF}}$ by the manifold optimization method. 
In the manifold optimization, the computational complexity is mainly rendered by the conjugate gradient descent method \cite{long2017coordinated}. Similar to the Euclidean unconstrained optimization, the number of iterations of the gradient descent method converging to the manifold gradient norm for satisfying the control threshold $\varepsilon $ can be quantified by  $\mathcal{O}\left(1/\varepsilon^2\right)$ for the worst case \cite{Absil}. For the conjugate gradient descent method on a complex circle manifold, the computational complexity of each iteration is characterized by ${\mathcal{O}} \left({N}^2_{t}{N}^{t}_{RF}{N_s} \right)$ \cite{jung2017low}.

\subsection{Energy Efficiency Analysis}

In this subsection, we analyze the energy efficiency of applying the HAD architecture and full-digital architecture. According to the description in \cite{yu2016alter,Aryan2021joint}, the energy efficiency at the transmitting side is defined as the ratio of spectral efficiency to power consumption, which is explicitly given by
\begin{equation}
{{R}_{p}}=\frac{{R}}{{P}_{sum}},
\end{equation}
where ${R}$ represents the spectral efficiency, and ${P}_{sum}$ is the total power consumption of the transmitter.
For the HAD architecture VBA-ISAC system, ${P}_{sum}$ is summed across
\begin{equation}\label{hy}
{{P}_{sum}}={{P}_{BB}}+{{N}^{t}_{RF}}{{P}_{RF}}+{N_t}{{P}_{PA}}+{{N}^{t}_{RF}}{N_t}{{P}_{PS}},
\end{equation}
where ${P}_{BB}$ represents the power consumption of the digital baseband in the transmitter; ${P}_{RF}$ represents the power consumption of each RF chain; ${P}_{PA}$ represents the power consumption of each linear amplifier; ${P}_{PS}$ represents the power consumption of each phase shifter. 
Similarly, for the full-digital architecture VBA-ISAC system, ${P}_{sum}$ is summed across
\begin{equation}\label{fully}
{{P}_{sum}}={{P}_{BB}}+{N_t}{{P}_{RF}}+{N_t}{{P}_{PA}}+{N_t}{{P}_{PS}}.
\end{equation}
In the full-digital structure, the number of RF chains is equal to the number of antennas ${N_t}$. However, that of hybrid structure is only ${{N}^{t}_{RF}}$. Since each RF chain in the hybrid structure is connected to all antennas, the number of phase shifters is thus ${N}^{t}_{RF} N_t$.

\section{Simulation Results} \label{V}
In this section, we show the numerical results to illustrate and discuss the superiority of our proposed VBA-ISAC beamforming design. Specifically, we first show the simulation results of predicting AoI, which is obtained by calculating the driving path of the vehicle. Then, we show the simulation results of beampatterns and communication performance, respectively. In the simulations, the number of transmit antennas for ISAC systems is $81$, where ULA is adopted.

\subsection{Sensing Performance}
\begin{figure}
  \centering
  \includegraphics[width=0.42\textwidth]{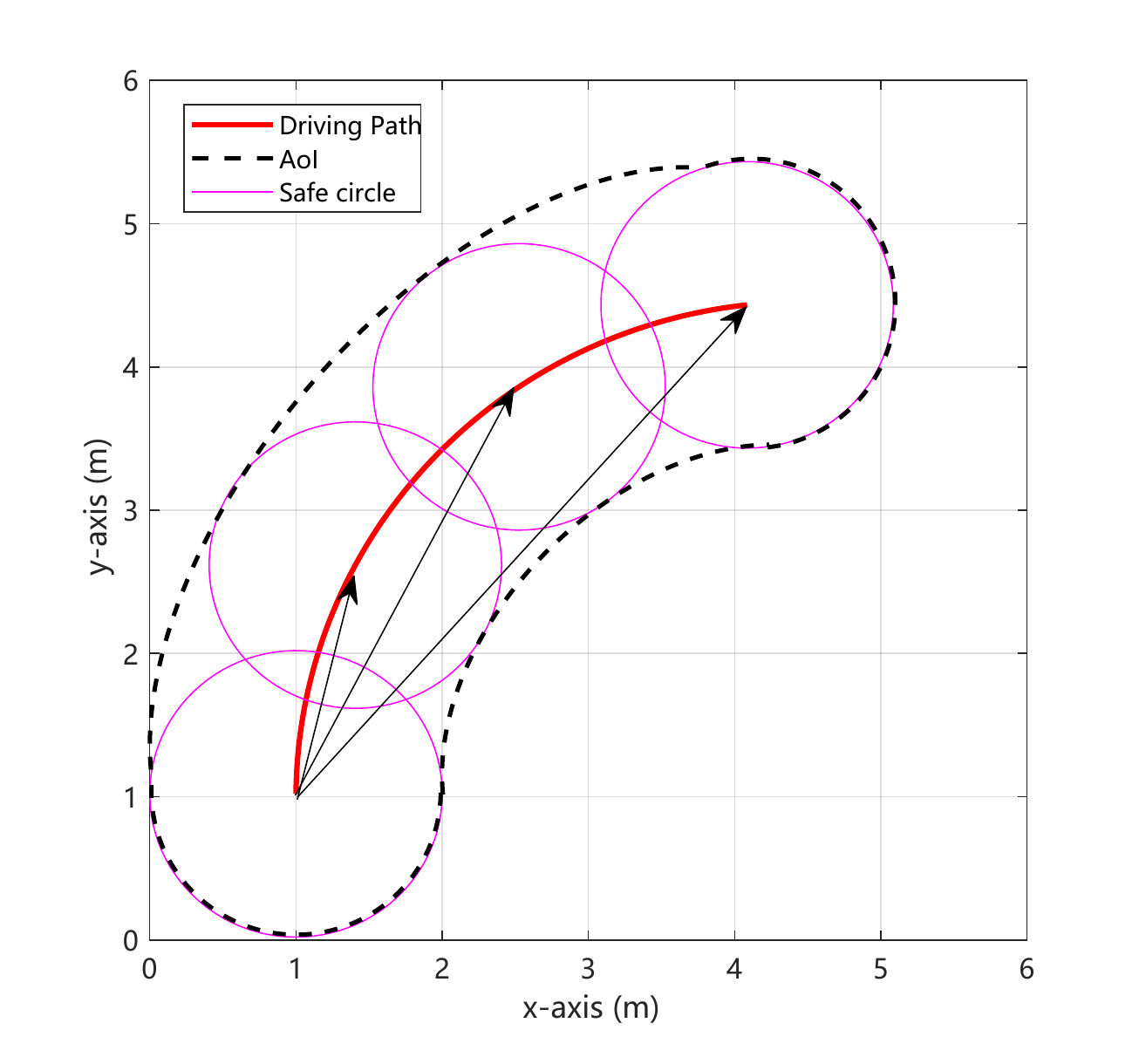}
  \caption{Simulation results of the driving path and the predicted AoI.} 
  \label{fig6}
\end{figure}

To carry out numerical simulations, we first need to set up parameters for the vehicle's kinematic model. Specifically, the acceleration of the vehicle is set to ${a} = $ 1~m/s$^2$; the steering angle and initial driving direction of the vehicle are set to ${\phi}={30}^{\circ}$ and ${\vartheta}={0}^{\circ}$; the initial position on the two-dimensional plane is set to $(1,1)$; the time slot is set to ${\Delta t}= $ 0.2~s; the initial velocity is set to $ v=$ 20~m/s; the distance between the front and rear wheels is set to ${l} = $ 2~m; the radius of the vehicle safety zone is set to ${r}_s = $ 1~m. The simulation parameters of the vehicle's kinematic model are summarized in Table I. Based on the simulation setups given above, the simulation results of the driving path of the vehicle on the two-dimensional plane and the predicted AoI are shown in Fig.~\ref{fig6}. In Fig.~\ref{fig6}, the red line is the driving path of the vehicle, and the circle represents the safety zone. For analytical simplicity, the driving path of the vehicle within ${\Delta t}$ is evenly divided into three stages. There are accordingly three vehicle positions: $(1.387, 2.581)$, $(2.526, 3.861)$, and $(4.085, 4.433)$ with corresponding AoI.

\begin{table}
\renewcommand{\arraystretch}{}
\caption{Simulation Parameters for the Vehicle's Kinematic Model}
\centering
\begin{tabular}{lr}
\hline
\bf Simulation parameters & \bf Number \\
\hline
Acceleration ${a}$ & 1~m/s$^2$  \\
Steering angle ${\phi}$ & ${30}^{\circ}$ \\
Initial driving direction ${\vartheta }$ & ${0}^{\circ}$   \\
Initial position & $(1,1)$   \\
Time slot ${\Delta t}$ & 0.2~s   \\
Initial velocity ${v}$ & 20~m/s   \\
Length of front and rear wheels ${l}$ & 2~m   \\
Radius of safety zone ${r}_s$ & 1~m   \\
\hline
\end{tabular}
\end{table}

\begin{figure}
  \centering
  \includegraphics[width=0.5\textwidth]{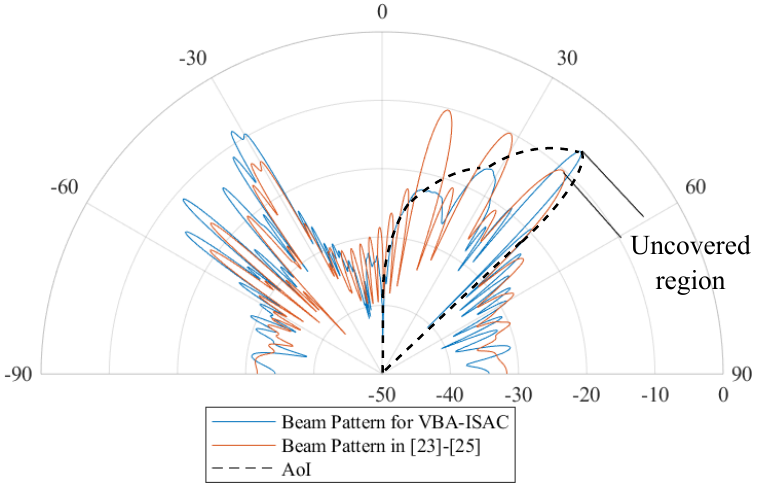}
  \caption{Simulation results of the beampatterns by our proposed method and the benchmarks.} 
  \label{fig7}
\end{figure}

The parameters of the sensing model are determined by the AoI. From the simulation results of AoI of the vehicle, the number of radar pointing angles is set to ${K}=3$. According to (\ref{theta}), pointing angles ${\theta}_k$  can be calculated as ${\theta}_1=14.1^{\circ}$, ${\theta}_2=28.1^{\circ}$, and ${\theta}_3=41.9^{\circ}$, respectively. According to (\ref{dsk}), the sensing distance at pointing angle ${\theta}_k$ can be calculated as ${d}^{s}_1=$2.7~m, ${d}^{s}_2=$4.2~m, and ${d}^{s}_3=$5.6~m. Numbers of antennas ${N}_k$ to form narrow beams are allocated as ${N}_1:{N}_2:{N}_3=2.7^4:4.2^4:5.6^4\approx4:18:59$. 
Based on the simulation setups given above, the simulation results of the beampatterns are shown in Fig.~\ref{fig7}. In \cite{liu2018toward,liu2019hybrid,dong2022vpaitmin}, the radar pointing angles depend on the entire area required to be scanned. And the number of antennas to form a desired beam at each pointing angle is evenly distributed. From Fig.~\ref{fig7}, it can be hard to accurately cover the AoI.
Compared with these intuitive benchmarks, setting radar pointing angle ${\theta}_k$ and the number of antennas ${N}_k$ based on AoI for a vehicle is more directional and accurate to cover the AoI, as the power of the radar can be more concentrated on the target area.

\begin{figure}
  \centering
  \includegraphics[width=0.48\textwidth]{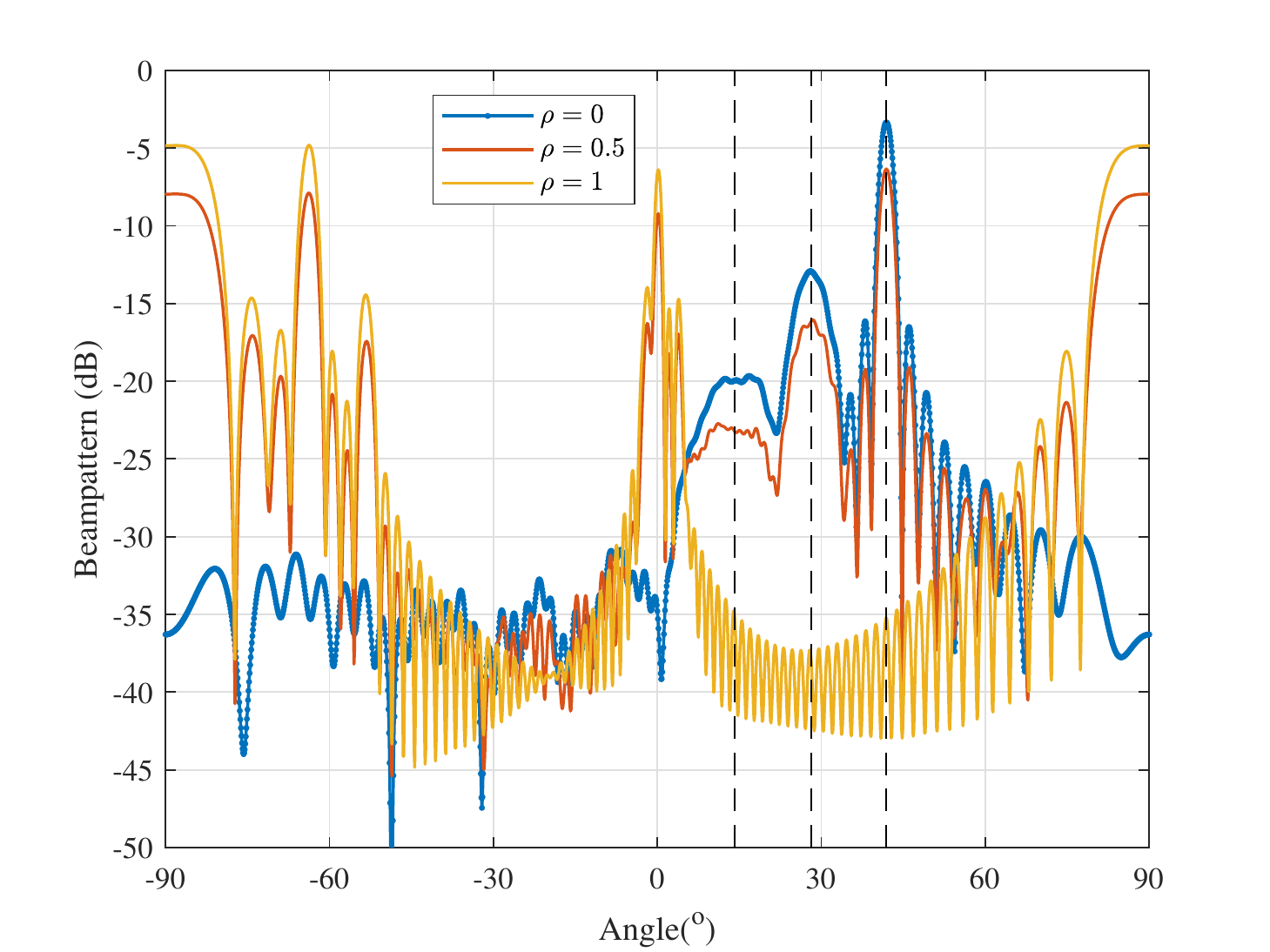}
  \caption{Influence on the sensing performance with different trade-off parameters $\rho$ = 0, 0.5 or 1.} 
  \label{fig8}
\end{figure}

Besides, we investigate the impact of trade-off parameter $\rho$ on the sensing performance. The simulation results of beampatterns with $\rho=0$, $0.5$ and $1$ are shown in Fig.~\ref{fig8}. In an extreme case when $\rho=0$, the VBA-ISAC system becomes a radar-only system, where the communication performance is not considered. In this case, we obtain an optimal beampattern.
It can also be seen that when trade-off $\rho$ is smaller, the beampattern is closer to the optimal beampattern. That is because the smaller the trade-off parameter is, the greater weight of sensing in the objective function of (\ref{model1}) will be. Therefore, the more main lobe of the beam is used to sense the radar pointing angles.  On the other hand, when $\rho$ is close to 1, the beampattern becomes rather poor and can hardly satisfy the sensing demand, since the power for sensing is rarely concentrated on the AoI. In this case, the VBA-ISAC system becomes a communication-only system.

\subsection{Communication Performance}

\begin{figure}
  \centering
  \includegraphics[width=0.48\textwidth]{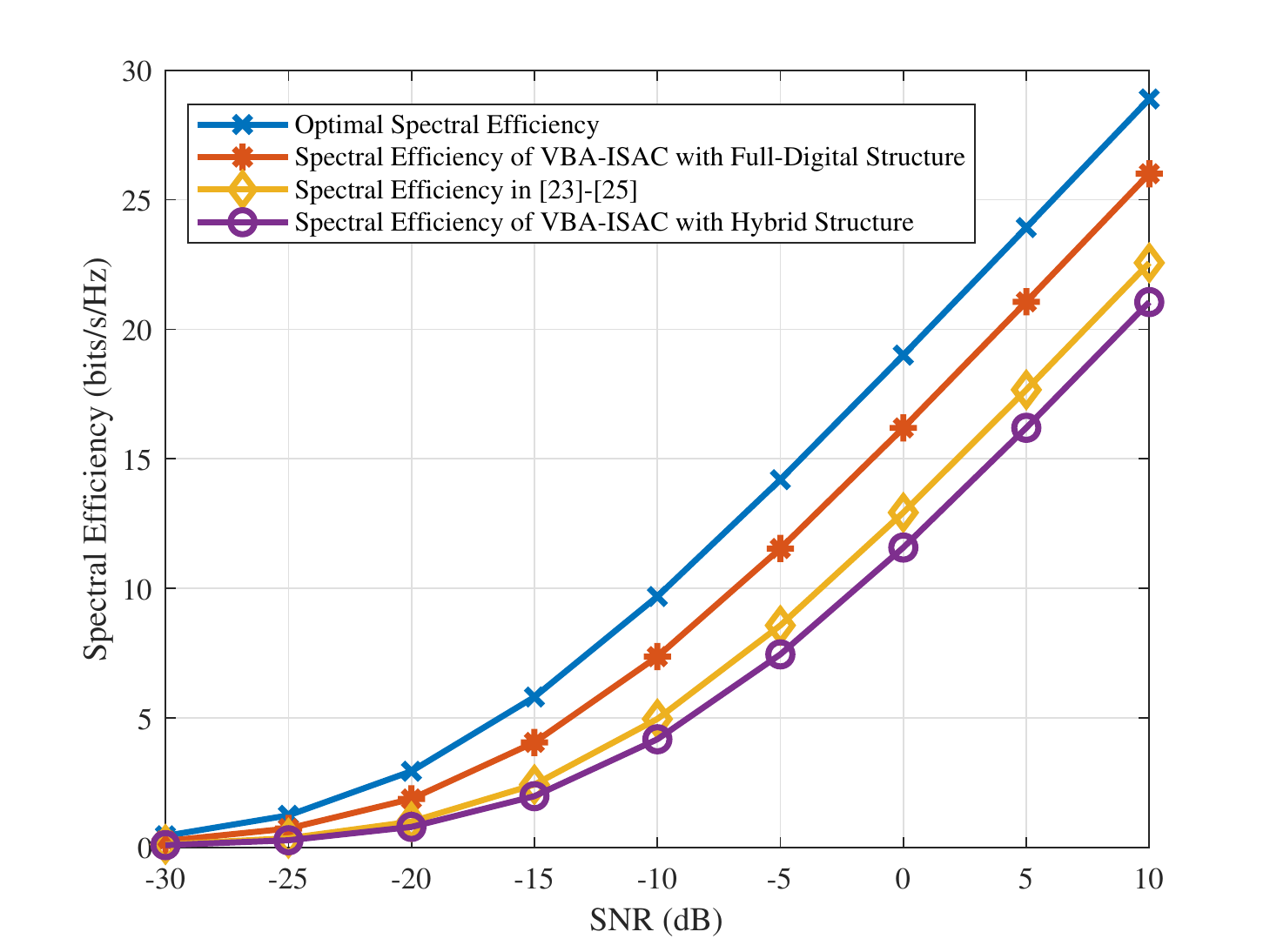}
  \caption{Simulation results of spectral efficiency for VBA-ISAC and the benchmarks.} 
  \label{fig9}
\end{figure}

In the simulation, the parameter setup is as follows. The number of receive antennas at the communication receiver ${N}_{r}=16$, where ULA is adopted. The number of transmission paths in the channel $L=10$, and the channel power gain ${\alpha _l}$ of each path is with deviation $\sigma_{\alpha}^2=1$. Departure angle ${\theta _{t,l}}$ and arrival angle ${\theta _{r,l}}$ are uniformly distributed within $[-90^{\circ},90^{\circ}]$. The number of data streams is set to ${N}_{s} = 3$. 
The communication performance of VBA-ISAC in spectral efficiency is demonstrated in Fig.~\ref{fig9}, where the trade-off factor $\rho=0.5$.
 It can be seen from Fig.~\ref{fig9} that the spectral efficiency of VBA-ISAC is lower than that of the communication-only system. That is because part of the main beam of the VBA-ISAC system is used to sense the AoI. As a result, the main lobe of the beam focusing on communication is reduced, leading to an inevitable performance loss. 
Fortunately, the spectral efficiency of VBA-ISAC is sufficiently close to the optimal spectral efficiency.  Moreover, the spectral efficiency of VBA-ISAC is significantly higher than the spectral efficiency in \cite{liu2018toward,liu2019hybrid,dong2022vpaitmin}. This is because, more beam power is required in \cite{liu2018toward,liu2019hybrid,dong2022vpaitmin} to cover the AoI. As a result, the beam power used for communication in \cite{liu2018toward,liu2019hybrid,dong2022vpaitmin} is reduced, resulting in a decrease in spectral efficiency. This phenomenon also validates the superiority of the proposed VBA-ISAC beamforming design scheme.

\begin{figure}
  \centering
  \includegraphics[width=0.48\textwidth]{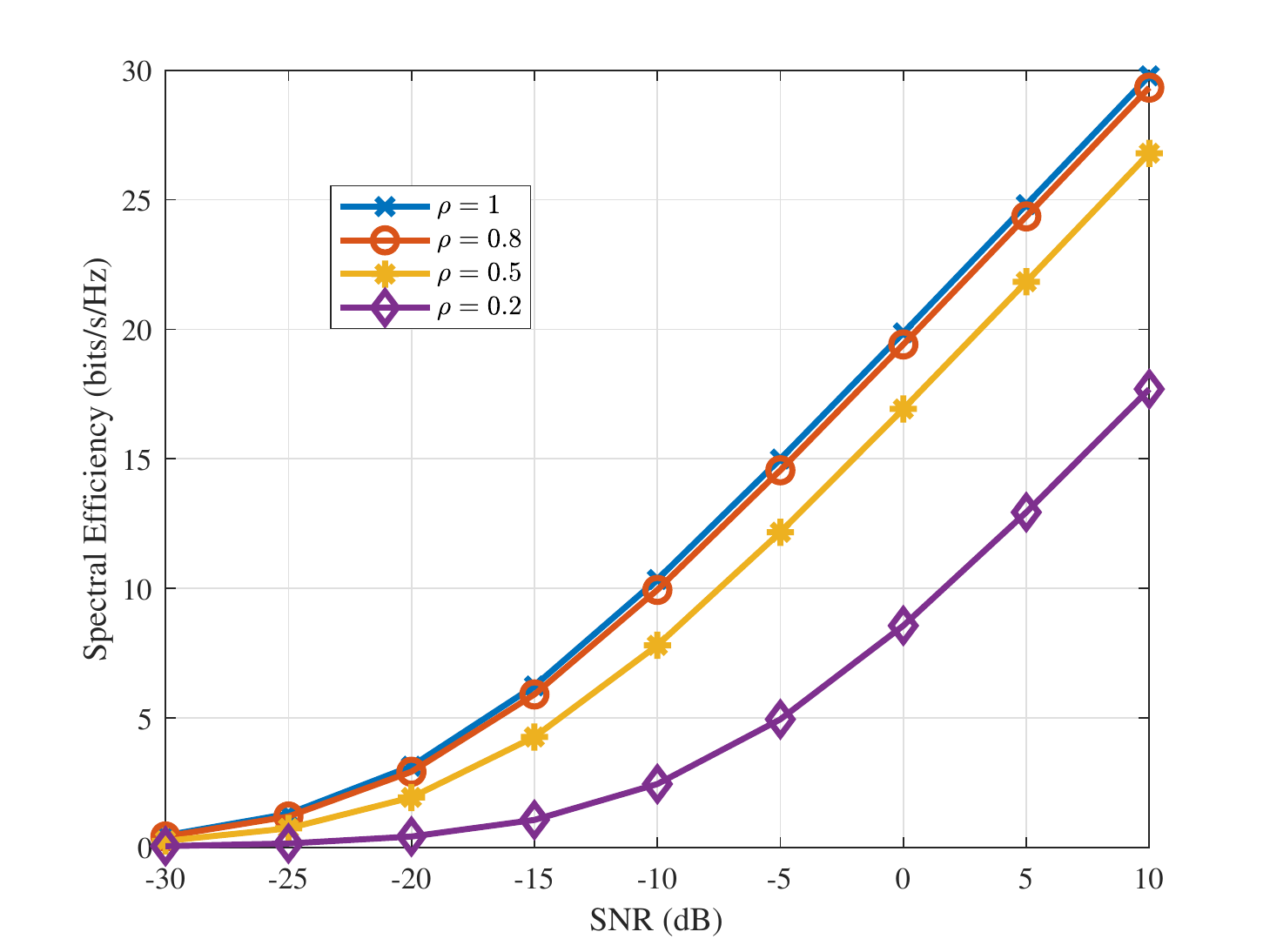}
  \caption{Spectral efficiency versus trade-off parameter $\rho$.} 
  \label{fig10}
\end{figure}

 The simulation results of spectral efficiency with $\rho=0.2$, $0.5$, $0.8$ and $1$ are shown in Fig.~\ref{fig10}. It is observed that as $\rho$ gradually increases, the spectral efficiency increases. This observed trend is aligned with the expectation because the main lobe of the beam used for communication increases as $\rho$ increases.
Combining the simulation results in Fig.~\ref{fig8} and Fig.~\ref{fig10}, one can conclude that the communication and sensing performance of the ISAC system can be modified by adjusting the trade-off factor $\rho$. In practical use, $\rho$ is determined by the functional requirements of users. If the users wish to achieve better communication performance, a larger $\rho$ can be set. Conversely, if the users wish to achieve better sensing performance, a smaller $\rho$ can be set. Thus, the best setting of $\rho$ depends on the  user's quality of service requirements.

\begin{figure}
  \centering
  \includegraphics[width=0.48\textwidth]{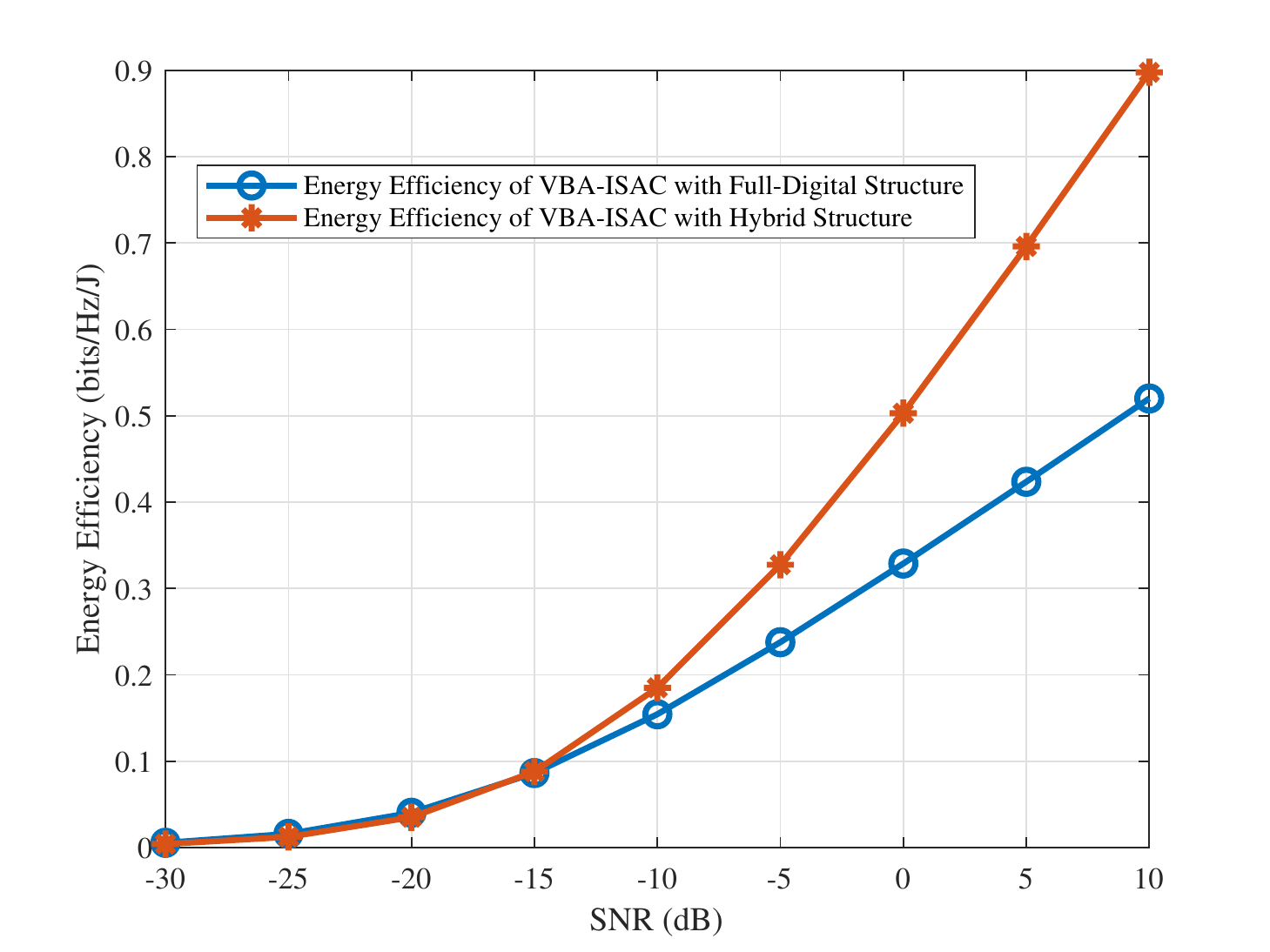}
  \caption{Energy efficiency of VBA-ISAC with full-digital structure and hybrid structure, given $\rho$ = 0.5.} 
  \label{fig12}
\end{figure}

For comprehensiveness, we also compare the performance of VBA-ISAC with fully digital and HAD structures in terms of  energy efficiency. In the simulations, the power consumption parameters are set as follows: ${P}_{BB} = 10 $ W, ${P}_{RF} = 300 $ mW, ${P}_{PA} = 100 $ mW, and ${P}_{PS} = 10 $ mW. The number of RF chains of the transmitter is set to ${N}^{t}_{RF} = 3$. The simulation results of the spectral efficiency VBA-ISAC system with the full-digital and hybrid RF structures are shown in Fig.~\ref{fig9}. It is observed that the spectral efficiency of the hybrid structure is lower than that of a full-digital structure. The simulation results of energy efficiency are shown in Fig.~\ref{fig12}, where $\rho=0.5$. It can be clearly seen from Fig.~\ref{fig12} that the energy efficiency of VBA-ISAC with the hybrid structure is higher than that of the full-digital structure.

\subsection{Performance Over Time-Varying Channels}

\begin{figure}
  \centering
  \includegraphics[width=0.48\textwidth]{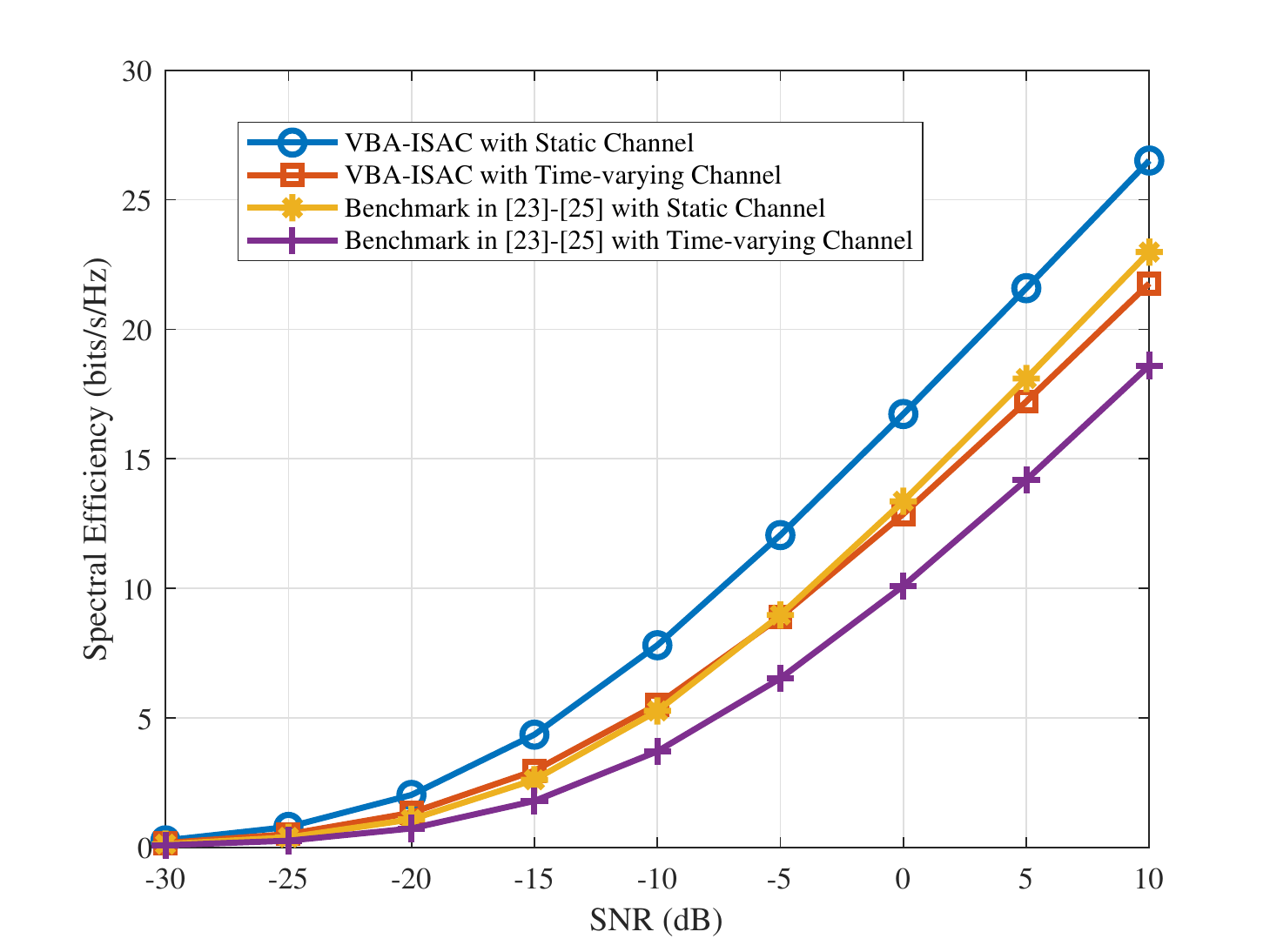}
  \caption{The simulation results of spectral efficiency with time-varying channel and static channel.} 
  \label{fig13}
\end{figure}

In (\ref{h}), we mainly consider a quasi-static mmWave channel model during a coherent time. To evaluate the impact of time-varying channels on the proposed scheme, we further conduct more simulations. Specifically, we split the time-varying channel into a static part and a time-varying part, and write the mmWave channel model in time-varying scenarios as
\begin{equation}
{\mathbf{H}_d} = {\mathbf{H}}+{\mathbf{H}_e},
\nonumber 
\end{equation}
where ${\mathbf{H}}$ stands for static channels shown in (\ref{h}); ${\mathbf{H}_e}$ represents time-varying part. For ISAC beamforming designs, ${\mathbf{H}_e}$ can also be regarded as the channel estimation errors caused by Doppler shift in practical time-varying scenarios. Without loss of generality, we assume each entry of ${\mathbf{H}_e}$ obeying zero mean and variance ${\sigma}_e$ complex Gaussian distribution \cite{cheng2018unified}. Simulation results are demonstrated in Fig.~\ref{fig13}. It is shown that the spectral efficiency of all schemes decreases when the time-varying part is not known.
Under the same level of unknown time-varying part, the spectral efficiency of the proposed VBA-ISAC scheme is still higher than the benchmarks in \cite{liu2018toward,liu2019hybrid,dong2022vpaitmin}.

\section{Conclusion} \label{VI}
The communication and sensing modules of traditional vehicle-mounted equipment are placed in isolation, resulting in low utilization of wireless  spectrum and hardware resources. To address this problem, we proposed a VBA-ISAC beamforming design for vehicle-mounted transmitters. By the proposed design, we predicted the trajectory based on the behavior of the vehicle. By introducing a safe zone, the AoI was determined according to the predicted driving path. After selecting the interesting pointing angles in the AoI, a desired radar beamformer was devised. Simultaneously, the vehicular transmitter was also able to communicate with the RSU. Then, we formulated the VBA-ISAC beamforming design as an optimization problem and introduced a trade-off factor to balance the communication and sensing performance. A tailored SDR algorithm was proposed to solve the formulated optimization problem. To cope with the large power consumption and high cost of VBA-ISAC system with full-digital architecture, we proposed and analyzed the VBA-ISAC beamforming design with the hybrid architecture. The numerical results demonstrated that the proposed beamforming design outperformed the benchmarks in both spectral efficiency and radar beampattern.

\bibliographystyle{IEEEtran} 
\bibliography{IEEEabrv,bib}

\end{document}